\documentclass[prd,preprintnumbers,amsmath,amssymb,floatfix]{revtex4}
\usepackage{amsmath,amsfonts,latexsym,amssymb,graphicx,graphics,epsfig,subfigure,color,makeidx}
\usepackage{xcolor,diagbox}
\usepackage{multirow}
\usepackage[colorlinks,linkcolor=blue,anchorcolor=blue,citecolor=green,urlcolor=blue]{hyperref}
\usepackage{mathrsfs}

\begin{document}
\title{Quasinormal ringing of thick braneworlds with a finite extra dimension}

\author{Hai-Long Jia$^{a}$$^{b}$\footnote{220220939521@lzu.edu.cn}}
\author{Wen-Di Guo$^{a}$$^{b}$\footnote{guowd@lzu.edu.cn}}
\author{Qin Tan$^{c}$\footnote{tanqin@hunnu.edu.cn}}
\author{Yu-Xiao Liu$^{a}$$^{b}$\footnote{liuyx@lzu.edu.cn, corresponding author}}

\affiliation{
$^{a}$Lanzhou Center for Theoretical Physics, 
    Key Laboratory for Quantum Theory and Applications of the Ministry of Education,
    Key Laboratory of Theoretical Physics of Gansu Province,
    School of Physical Science and Technology,
    Lanzhou University, Lanzhou 730000, China \\
$^{b}$Institute of Theoretical Physics \& Research Center of Gravitation,
    Lanzhou University, Lanzhou 730000, China \\
$^{c}$Department of Physics,
    Key Laboratory of Low Dimensional Quantum Structures and Quantum Control of Ministry of Education,
    Synergetic Innovation Center for Quantum Effects and Applications,
    Hunan Normal University,
    Changsha, 410081, Hunan, China
}

\begin{abstract}
    In this work, we investigate the quasinormal modes of the Poincaré thick brane with a finite extra dimension. 
    Unlike the case with an infinite extra dimension, 
    the gravitational effective potential exhibits three distinct shapes within different ranges of the parameter $n$ 
    in the warp factor: harmonic oscillator potential, Pöschl-Teller potential, and volcano-like potential. 
    We then study various types of perturbations in this system. 
    Utilizing a combination of analytical, semi-analytical, and numerical methods, 
    we obtain the quasinormal modes of the perturbed fields. 
    Our findings reveal a set of discrete quasinormal modes for the thick brane, similar to those of black holes. 
    Interestingly, when $n=1$, the quasinormal modes exhibit purely imaginary behavior. 
    This study may provide a new way to detect the existence of extra dimensions. 
\end{abstract}

\maketitle
%%%%%%%%%%%%%%%%%%%%%%%%%%%%%%%%%%%%%%%%%%%%%%%%%%%%%%%%%%%%%%%%%%%%%%%%%%%%%%%%%%%%%%%%%%%%%%%%%%%%%%%%%%%%%%%%%%%%%%%%%%%%%
\section{Introduction} \label{Introduction}

The detection of gravitational waves in 2015 by the Laser Interferometer Gravitational-Wave Observatory (LIGO) and Virgo~\cite{LIGOScientific:2016aoc} 
sparked renewed interest in black hole physics, particularly in the realm of quasinormal modes (QNMs)~\cite{Berti:2009kk,Cardoso:2016rao,Kokkotas:1999bd,Konoplya:2011qq,Jusufi:2020odz,Nollert:1999ji,Cheung:2021bol}. 
In general, QNMs refer to the characteristic modes of a dissipative system. 
It is well known that QNMs can reveal the characteristic information of a black hole, 
such as mass, charge, angular momentum, and so on. Additionally, QNMs play a crucial role in other fields of physics.
For example, as mentioned in Ref.~\cite{Berti:2009kk}, QNMs have become a valuable tool for detecting extra-dimensional signals. 
There are two main approaches: investigating the QNMs of black holes on the branes~\cite{Chakraborty:2017qve,Prasobh:2014zea,Dey:2020pth,Hashemi:2019jlt,Chen:2016qii,Konoplya:2003dd,Cardoso:2003vt,Seahra:2004fg,Kanti:2005xa,Seahra:2006tm,Kanti:2006ua,Ishihara:2008re,Chung:2015mna,Bronnikov:2019sbx,Dey:2020lhq,Banerjee:2021aln,Mishra:2021waw,Lin:2022hus,Bohra:2023vls,Zinhailo:2024jzt}, 
or directly researching the QNMs of brane models~\cite{Seahra:2005wk,Seahra:2005iq,Tan:2022vfe,Tan:2023cra}. 

The braneworld concept first appeared in string theory~\cite{Polchinski:1995mt}, 
and was later applied to solve the hierarchical problem between the Electroweak and Planck scales~\cite{Antoniadis:1990ew,Arkani-Hamed:1998jmv,Antoniadis:1998ig,Randall:1999ee}.
Its basic idea is that, gravitons can propagate throughout the entire spacetime (bulk), but all other fundamental particles are restricted on the brane. 
Since the 1990s, the most popular braneworld model has been the warped extra-dimensional model proposed by Randall and Sundrum (RS)~\cite{Randall:1999ee,Randall:1999vf}. 
There are two thin branes (RS-\uppercase\expandafter{\romannumeral1} model) or one brane (RS-\uppercase\expandafter{\romannumeral2} model) embedded in a five-dimensional Anti-de Sitter (AdS) spacetime. 
In the RS-\uppercase\expandafter{\romannumeral2} model, it was found that the four-dimensional Newtonian potential at a large scale can be recovered on the brane 
because of the localization of the graviton zero mode even if the extra dimension is infinite, 
while the potential at a small scale is corrected due to the massive Kaluza-Klein (KK) spectrum~\cite{Randall:1999vf}. 
Currently, since the experimental observation that Newton's inverse-square law still holds 
at the submillimeter scale~\cite{Tu:2007zz,Yang:2012zzb,Tan:2016vwu,Tan:2020vpf,Lee:2020zjt,Du:2022veu}, 
this correction has not yet been confirmed.
On the other hand, a characteristic of the thin brane model is that the energy density of the brane is a delta function along the extra dimension, ignoring the intrinsic structure of the brane. 
Therefore, based on the domain wall model~\cite{Akama:1982jy,Rubakov:1983bb} and RS-\uppercase\expandafter{\romannumeral2} model~\cite{Randall:1999vf},
thick brane models emerged as a seamless extension offering novel possibilities~\cite{DeWolfe:1999cp,Gremm:1999pj,Csaki:2000fc}. 
In the past two decades, various solutions of thick brane models have been found~\cite{DeWolfe:1999cp,Gremm:1999pj,Csaki:2000fc,Wang:2002pka,Dzhunushaliev:2010fqo,Dzhunushaliev:2011mm,Guo:2011wr,Liu:2012gv,German:2012rv}, 
and the localizations of gravity and various matter fields on the brane have been discussed~\cite{Gregory:2000jc,Melfo:2006hh,Almeida:2009jc,Liu:2009ve,Liu:2009dw,Liu:2009uca,Zhao:2009ja,Liu:2011wi,Xie:2015dva,Gu:2016nyo,Zhong:2016iko,Zhou:2017xaq,Xie:2021ayr,Xu:2022xxd}.
For more information on thick brane models, please refer to these review articles~\cite{Liu:2017gcn,Dzhunushaliev:2009va,Ahluwalia:2022ttu}.

As early as 2005, Seahra found that in the RS-\uppercase\expandafter{\romannumeral2} model, for a bulk observer, there are metastable gravitational wave bound states (also called resonance states), 
while for an observer on the brane, these states can be viewed as a set of discrete QNM spectrum of decaying massive gravitons. 
Recently, some of us found that, similar to the thin brane case, a thick brane can also possess a series of separated QNMs~\cite{Tan:2022vfe,Tan:2024url}.
Furthermore, we also discovered that those long-lived QNMs have a close relationship with the resonance modes of a thick brane~\cite{Tan:2023cra}. 
This paper aims to investigate the difference in QNMs between the thick brane with a finite extra dimension and that with an infinite extra dimension. 
We also discuss the stability of the brane under various types of perturbations.

In this paper, we use capital Latin letters $M,N,\dots = 0,1,2,3,5$, Greek letters $\mu,\nu,\dots=0,1,2,3$, and Latin letters $i,j,\dots=1,2,3$ 
to label the five-dimensional spacetime, the four-dimensional spacetime, and the three-dimensional space coordinates, respectively. 
This paper is organized as follows:    
In Sec.~\ref{Model}, we provide a review of a solution for the thick brane and the linear perturbation of the background. 
In Sec.~\ref{matter fields}, we review the effective potential of bulk test matter fields with spins 0 and 1 on this thick brane.
In Sec.~\ref{Quasinormal modes}, we investigate the QNMs of the tensor perturbation and scalar perturbation of the background, 
as well as the bulk matter fields, using analytical, semi-analytical, and numerical methods. 
Finally, in Sec.~\ref{conclusion}, we present our conclusions. 

%%%%%%%%%%%%%%%%%%%%%%%%%%%%%%%%%%%%%%%%%%%%%%%%%%%%%%%%%%%%%%%%%%%%%%%%%%%%%%%%%%%%%%%%%%%%%%%%%%%%%%%%%%%%%%%%%%%%%%%%%%%%%
\section{Review of thick braneworld model} \label{Model}

\begin{figure*}[htb]
    \begin{center}
    \subfigure[~Bulk scalar field]  {\label{phiy-Plot}
    \includegraphics[width=5.6cm]{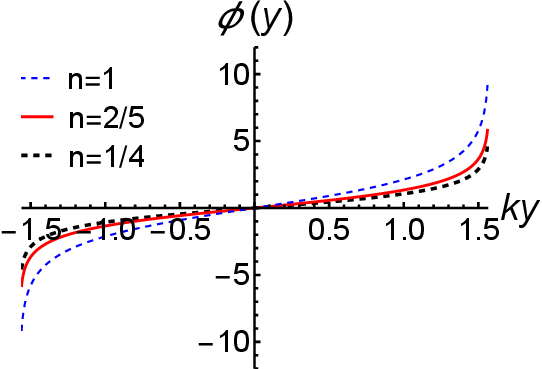}}
    \subfigure[~Scalar potential]  {\label{Vphiy-Plot}
    \includegraphics[width=5.6cm]{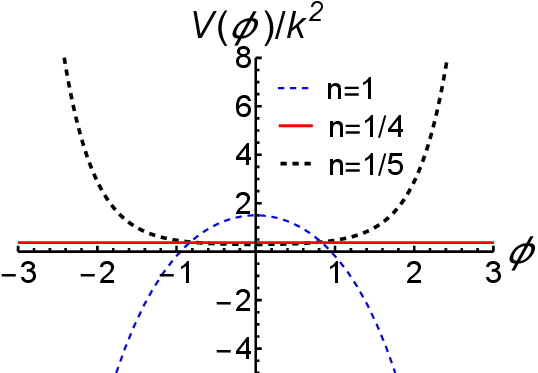}}
    \subfigure[~Curvature scalar]  {\label{Ry-Plot}
    \includegraphics[width=5.6cm]{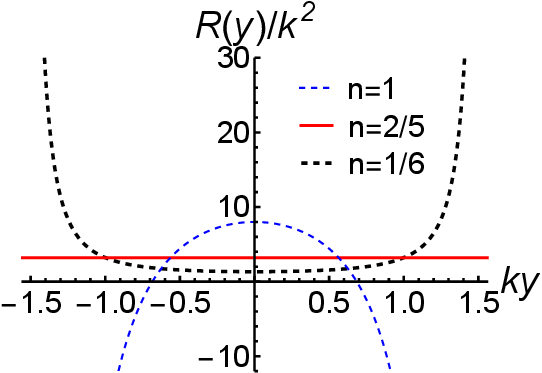}}
    \end{center}
    \caption{The shapes of the bulk scalar field~\eqref{phiy}, the scalar potential~\eqref{Vphi}, and the curvature scalar~\eqref{Ry}.}
    \label{Figure-1}
\end{figure*}

In this section, we first review the model of the thick brane with a finite extra dimension, generated by a bulk scalar field $\phi$. 
The action of this thick brane model is~\cite{DeWolfe:1999cp,Csaki:2000fc}
\begin{equation}
    \label{action}
    S_{5}=\int d^4x dy \sqrt{-g} \left( \frac{\hat{M}^3}{2}R - \frac{1}{2}g^{MN}\nabla_{M}\phi \nabla_{N}\phi - V(\phi) \right) , 
\end{equation}
where $\hat{M}$ is the five-dimensional Planck mass scale and $y$ represents the coordinate of the extra dimension. 
Here, we set $\hat{M}=1$ for convenience.  
By varying the action~\eqref{action} with respect to the scalar field and the metric, we can obtain 
\begin{align}
    \label{phi-field}
    g^{MN}\nabla_{M}\nabla_{N}\phi &= \frac{\partial V(\phi)}{\partial \phi}, \\
    R_{MN}-\frac{1}{2}g_{MN}R &= g_{MN} \left(- \frac{1}{2}g^{PQ}\nabla_{P}\phi \nabla_{Q}\phi - V(\phi) \right) + \nabla_{M}\phi \nabla_{N}\phi . 
    \label{E-field}
\end{align}
We only consider the static flat brane that satisfies four-dimensional Poincaré symmetry.  
The metric is~\cite{Randall:1999ee}
\begin{equation}
    \label{line-element} 
    ds_5^2=e^{2A(y)}\eta_{\mu \nu}dx^{\mu} dx^{\nu}+dy^2,
\end{equation}
where $e^{2A(y)}$ is the warp factor and $\eta_{\mu\nu}=\text{diag}(-1,1,1,1)$ is the four-dimensional Minkowski metric. 
By assuming $\phi=\phi(y)$ and substituting Eq.~\eqref{line-element} into Eqs.~\eqref{phi-field} and~\eqref{E-field}, we obtain the specific equations of motion as follows:
\begin{align}
    \label{phi-equation}
    \phi''+4 A' \phi' &= \frac{\partial V(\phi)}{\partial \phi} ,\\
    \label{E-equation-1}
    6 A'^2 + 3A'' &= -\frac{1}{2} \phi'^2 - V(\phi) , \\
    \label{E-equation-2}
    6 A'^2 &= \frac{1}{2} \phi'^2 - V(\phi) ,     
\end{align}
where the prime denotes the derivative with respect to the extra coordinate $y$.
Through the equations mentioned above, we can solve for the expressions of $\phi(y)$ and $V(y)$ in terms of $A(y)$ as follows: 
\begin{align} 
    \label{phiy-expression}
    \phi(y) &= \int \sqrt{-3 A''(y)} \,dy ,\\
    \label{V-expression}
    V(\phi(y)) &= -\left( 6 A'^2 + \frac{3}{2}A'' \right) .
\end{align}
It can be demonstrated that Eq.~\eqref{phi-equation} can be derived from Eqs.~\eqref{E-equation-1} and~\eqref{E-equation-2}; 
therefore, only two of the three equations \eqref{phi-equation}, \eqref{E-equation-1}, and \eqref{E-equation-2} are independent. 
However, since there are three functions: $A(y)$, $\phi(y)$, and $V(\phi)$, we must specify one of these functions to solve the equations. 
These solutions can be obtained using the superpotential method~\cite{DeWolfe:1999cp,Bazeia:2008zx,Zhong:2013xga} or by fixing one of the variables. 
Here, we quote the results from Refs.~\cite{Gremm:2000dj,Barbosa-Cendejas:2007ucz}, whose warp factor is 
\begin{equation}
    \label{Ay-expression}
    A(y) = n \log\left(\cos(ky)\right) , 
\end{equation}
where $n$ and $k$ are both constants. For $n \leqslant 0$, the warp factor is either nonphysical or trivial. 
Thus, we set $n>0$, and $k$ characterizes the width of the warp factor and represents the characteristic energy scale within the bulk spacetime.
The range of the extra dimension $y$ is $-\frac{\pi}{2} \leqslant ky \leqslant \frac{\pi}{2} $.  
By substituting Eq.~\eqref{Ay-expression} into Eqs.~\eqref{phiy-expression} and~\eqref{V-expression}, 
we can obtain the solution: 
\begin{align}
    \label{phiy}
    \phi(y) &= \sqrt{3n} \log\left[ \sec(ky) + \tan(ky)\right] ,\\
    \label{Vphi}
    V(\phi) &= \frac{3k^2 n}{2} \left[1 + (1-4n)\sinh^2\left(\frac{\phi}{\sqrt{3n}}\right) \right]. 
\end{align}
Furthermore, we can obtain the explicit form of the five-dimensional curvature scalar: 
\begin{equation}
    \label{Ry}
    R(y) = 4n k^2 \left(5n +(2-5n) \sec^2 (ky)\right). 
\end{equation}
Plots of the bulk scalar field $\phi(y)$, the scalar potential $V(\phi)$, and the five-dimensional curvature scalar $R(y)$ are shown in Fig.~\ref{Figure-1}. 

Combining Eq.~\eqref{Vphi} and Fig.~\ref{Vphiy-Plot}, we can observe that 
the existence of a critical value $n=\tfrac{1}{4}$ concerning the shape of the scalar potential. 
When $n<\tfrac{1}{4}$, the potential well opens upward; conversely, when $n>\tfrac{1}{4}$, it opens downward.
From Fig.~\ref{Ry-Plot}, it is evident that there are naked singularities at the boundaries of the extra dimension, i.e., at $y=\pm \frac{\pi}{2k}$. 
These singularities could be addressed by lifting the five-dimensional geometry to ten dimensions or by employing string theory~\cite{Gremm:2000dj}. 
Gremm speculated that within the AdS/CFT correspondence, the singular space corresponds to a nonconformal theory such as the super Yang-Mills theory or quantum chromodynamics~\cite{Gremm:2000dj}. 
Notably, for the thick brane with an infinite extra dimension, such naked singularities do not occur, 
and the bulk scalar field exhibits a standard kink solution. 

To facilitate the study of the perturbations of the thick brane and bulk matter fields, 
it is convenient to transform the extra-dimensional coordinate $y$ into the conformal flat coordinate $z$, 
which involves the following coordinate transformation: 
\begin{equation}
    \label{transformation}
    dz=e^{-A(y)}dy. 
\end{equation}
By integrating both sides, we find that $z(y)=\int e^{-A(y)}dy$. This allows us to derive the analytical expressions for $z$ when $n$ is an integer: 
\begin{align}
    \label{z-y-1}
    &n=1:\quad  z= \frac{1}{k} \log \left( \frac{\cos(ky)}{1-\sin(ky)} \right) , \\  
    \label{z-y-2}
    &n=2:\quad  z= \frac{1}{k} \tan(ky) , \\  
    \label{z-y-3}
    &n=3:\quad  z= \frac{1}{2 k}\left( \operatorname{arctanh} (\sin(ky)) + \sec(k y)\tan(k y) \right) ,\\
    \label{z-y-4}
    &n=4:\quad  z= \frac{1}{k}\left( \tan(ky)+\frac{1}{3}\tan^3(ky) \right) ,\\
    &\dots  \nonumber
\end{align}
In fact, when $n\neq2m+1,m\in \mathbb{N}$, we have 
\begin{equation}
    \label{z-y-n}
    z(y) = \frac{\text{sgn}(ky)}{k} \left[ \frac{\sqrt{\pi}}{2}\frac{\Gamma(\frac{1-n}{2})}{\Gamma(1-\frac{n}{2})}+\frac{\cos(ky)^{(1-n)}}{n-1} \, {}_2\!F_1 \left(\frac{1}{2},\frac{1-n}{2},\frac{3-n}{2},\cos^2(ky)\right)\right],
\end{equation}
where $\text{sgn}(ky)$ is the sign function and ${}_2\!F_1\left(a,b,c,f(y)\right)$ is the hypergeometric function. 
It can be proven that when $n\geqslant 1$, the range of $z$ is $-\infty < z <\infty$; when $0<n<1$, this range is finite. 
Following the transformation \eqref{transformation}, the corresponding line-element~\eqref{line-element} becomes
\begin{equation}
    \label{line-element-z} 
    ds^2=e^{2A(z)}\left( \eta_{\mu \nu}dx^{\mu} dx^{\nu}+dz^2 \right) .
\end{equation}

In Sec.~\ref{The properties of effective potentials}, it can be shown that for the case of $0<n<1$, the corresponding potentials for the perturbations and bulk fields diverge at the boundary,
indicating that only normal modes exist in this case. 
Therefore, we will focus our discussion on the case of $n\geqslant 1$ for the problem of QNMs. 

%%%%%%%%%%%%%%%%%%%%%%%%%%%%%%%%%%%%%%%%%%%%%%%%%%%%%%%%%%%%%%%%%%%%%%%%%%%%%%%%%%%%%%%%%%%%%%%%%
\subsection{Tensor perturbation}

Now we consider the linear transverse-traceless tensor perturbation of the thick brane. 
In the conformal coordinate $z$, the perturbed metric reads 
\begin{equation}
    \label{line-element-conform} 
    ds^2=e^{2A(z)} \left[(\eta_{\mu \nu}+h_{\mu\nu})dx^{\mu} dx^{\nu}+dz^2 \right], 
\end{equation}
where $h_{\mu\nu}$ satisfies the transverse-traceless condition $\partial_{\alpha}h^{\alpha\beta}=\eta^{\alpha\beta}h_{\alpha\beta}=0$. 
Then, substituting the perturbed metric~\eqref{line-element-conform} into the field equation~\eqref{E-field}, 
we obtain the equation for the tensor perturbation: 
\begin{equation}
    \label{five-wave-equation}
    \left[\partial_{z}^{2} + 3 \left(\partial_{z} A(z)\right)\partial_{z} + \Box^{(4)}\right] h_{\mu\nu} = 0, 
\end{equation}
where $\Box^{(4)}=\eta^{\alpha\beta}\partial_{\alpha}\partial_{\beta}$ is the four-dimensional Minkowski D'Alembert operator. 
And one makes the ansatz~\cite{DeWolfe:1999cp}:  
\begin{equation}
    \label{ansatz}
    h_{\mu\nu}= \epsilon_{\mu\nu} e^{-\frac{3}{2}A(z)}e^{-ip_{j}x^{j}} \Psi_{gt}(t,z) , 
\end{equation}
where $\epsilon_{\mu\nu} = \text{constant}$.
In this paper, we use the subscripts $gt$ and $gs$ to denote the cases of the tensor and scalar perturbations of the thick brane, respectively, 
and $s$ and $v$ to denote the bulk scalar and bulk vector fields, respectively. 
Substituting the ansatz~\eqref{ansatz} into Eq.~\eqref{five-wave-equation}, 
we obtain a wave equation for $\Psi_{gt}(t,z)$:
\begin{equation}
    \label{gt-Phi-equation}
    \left[-\partial_{t}^{2} + \partial_{z}^{2} + V_{gt}(z) - p^2\right] \Psi_{gt}(t,z)=0 , 
\end{equation}
where $p^2=\delta^{ij}p_{i}p_{j}$, and
\begin{equation}
    \label{gt-effective-potential}
    V_{gt}(z)=\frac{3}{2}\partial_{z}^{2}A + \frac{9}{4}\left(\partial_{z}A\right)^2 
\end{equation}
is the effective potential for the tensor perturbation. 
Furthermore, by assuming $\Psi_{gt}(t,z) = e^{-i\omega t}\psi_{gt}(z)$, we can obtain a Schrödinger-like equation for the extra-dimensional part $\psi_{gt}(z)$: 
\begin{equation}
    \label{gt-Slike-equation}
    \left[-\partial_{z}^{2} + V_{gt}(z)\right] \psi_{gt}(z) = m^2 \psi_{gt}(z) ,
\end{equation}
where $m^2=\omega^2-p^2$ represents the mass of the KK mode of the tensor perturbation. In Ref.~\cite{DeWolfe:1999cp}, it was demonstrated that 
Eq.~\eqref{five-wave-equation} supports a massless zero mode and a series of massive KK modes.
The zero mode can be localized on the thick brane, and it satisfies $\psi_{gt}^{(0)}(z) \propto e^{\frac{3}{2}A(z)}$. 

%%%%%%%%%%%%%%%%%%%%%%%%%%%%%%%%%%%%%%%%%%%%%%%%%%%%%%%%%%%%%%%%%%%%%%%%%%%%%%%%%%%
\subsection{Scalar perturbation}

Next, let us consider the scalar perturbation of the thick brane. 
In this model, gravity is coupled with a scalar field. Therefore, when considering the scalar perturbations of the metric, we also need to consider the perturbation of the scalar field. 
In the longitudinal gauge~\cite{Giovannini:2001fh,Giovannini:2001xg,Kobayashi:2001jd}, the metric with scalar perturbations is
\begin{equation}
    \label{line-element-conform-scalar} 
    ds^2=e^{2A(z)} \left[ \left(1+ \varphi(x^{\mu},z) \right)\eta_{\mu \nu}dx^{\mu} dx^{\nu}+ \left( 1+ \xi(x^{\mu},z) \right) dz^2 \right] ,
\end{equation}
and the bulk scalar field is perturbed as
\begin{equation}
    \label{scalar-perturbation}
    \phi(x^{\mu},z)=\phi(z)+\delta\phi(x^{\mu},z) . 
\end{equation}
Here, $\varphi$, $\xi$, and $\delta\phi$ are all dynamical variables. 
By substituting the perturbed metric~\eqref{line-element-conform-scalar} and formula~\eqref{scalar-perturbation} into the field equations~\eqref{phi-field} and~\eqref{E-field}, 
we obtain the equations for the scalar perturbations: 
\begin{align}
    \label{zz-section}
    (z,z)&:\quad \Box^{(4)} \xi + 4 \partial_{z}^{2}\varphi-4 \partial_{z}A\partial_{z}\xi + 4\partial_{z}A\partial_{z}\varphi = -4\partial_{z}\phi \partial_{z}\delta\phi + \xi\left(2\partial_{z}^{2}A+6(\partial_{z}A)^2\right) -\frac{4}{3}e^{2A}\frac{\partial V(\phi)}{\partial \phi} \delta \phi , \\
    \label{muz-section}
    (\mu,z)&:\quad -\partial_{\mu}\partial_{z}\varphi + \partial_{z}A\partial_{\mu}\xi = \frac{2}{3}\partial_{z}\phi \partial_{\mu} \delta\phi , \\
    \label{munu-section}
    (\mu,\nu)&:\quad \partial_{\mu}\partial_{\nu}\left(\xi+2\varphi\right) + \eta_{\mu\nu} \left( \Box^{(4)} \varphi + \partial_{z}^{2}\varphi + 7\partial_{z}A\partial_{z}\varphi-2\xi\left(\partial_{z}^{2}A+3(\partial_{z}A)^2\right)-\partial_{z}\xi\partial_{z}A \right) =-\frac{4}{3}\eta_{\mu\nu}e^{2A} \frac{\partial V(\phi)}{\partial \phi} \delta \phi, \\
    \label{matter-section}
    \text{matter}&:\quad \Box^{(4)} \delta \phi+\partial_{z}^{2}\delta\phi +3\partial_{z}A\partial_{z}\delta\phi+\partial_{z}\phi\left(2\partial_{z}\varphi-\frac{1}{2}\partial_{z}\xi\right)-\xi\left(\partial_{z}^{2}\phi+3\partial_{z}A\partial_{z}\phi\right) = e^{2A} \frac{\partial^2 V(\phi)}{\partial \phi^2} \delta \phi. 
\end{align}
From Eq.~\eqref{muz-section} and the off-diagonal components of Eq.~\eqref{munu-section}, one can obtain the following constraints on these three scalar functions $\delta\phi$, $\varphi$, and $\xi$:  
\begin{align}
    \label{constraint1}
    \partial_{\mu}\partial_{\nu}\left(\xi+2\varphi\right) &= 0, \\
    \label{constraint2}
    2\partial_{z}\phi \, \delta\phi + 3 \left(\partial_{z}\varphi +2\partial_{z}A\,\varphi \right) &= 0 , 
\end{align}
which indicate that there is only one physical scalar degree of freedom. 
Then, substituting Eqs.~\eqref{constraint1} and~\eqref{constraint2} into Eqs.~\eqref{zz-section} and~\eqref{munu-section}, 
the main equation can be obtained as 
\begin{equation}
    \label{main-equation}
    \Box^{(4)}\varphi+\partial_{z}^{2}\varphi+\left(3\partial_{z}A-\frac{2\partial_{z}^{2}\phi}{\partial_{z}\phi}\right) \partial_{z}\varphi + \left(4\partial_{z}^{2}A-\frac{4\partial_{z}A \partial_{z}^{2}\phi}{\partial_{z}\phi}\right)\varphi =0.  
\end{equation}
By defining a new variable 
\begin{equation}
    \Phi(x^{\mu},z) = \frac{e^{-\frac{3}{2}A(z)}}{\partial_{z}\phi(z)}\varphi(x^{\mu},z) , 
\end{equation}
and substituting it into Eq.~\eqref{main-equation}, we obtain  
\begin{equation}
    -\partial_{z}^{2}\Phi + V_{gs}(z) \Phi = \Box^{(4)}\Phi , 
\end{equation}
where
\begin{equation}
    V_{gs}(z) = -\frac{5}{2}\partial_{z}^{2}A+\frac{9}{4}(\partial_{z}A)^2 -\frac{\partial_{z}^{3}\phi}{\partial_{z}\phi} + \partial_{z}A \frac{\partial_{z}^{2}\phi}{\partial_{z}\phi} + 2 \left(\frac{\partial_{z}^{2}\phi}{\partial_{z}\phi}\right)^2
\end{equation}
is the effective potential for the scalar perturbation $\Phi(x^{\mu},z)$.  
By defining $\Phi(x^{\mu},z) \propto e^{-ip_{j}x^{j}} \Psi_{gs}(t,z)$, we obtain a wave equation for $\Psi_{gs}(t,z)$ similar to Eq.~\eqref{gt-Phi-equation}: 
\begin{equation}
    \label{gs-Phi-equation}
    \left[-\partial_{t}^{2} + \partial_{z}^{2} + V_{gs}(z) - p^2\right] \Psi_{gs}(t,z)=0 .
\end{equation}
Finally, by assuming $\Psi_{gs}(t,z) = e^{-i\omega t}\psi_{gs}(z)$, we can obtain a Schrödinger-like equation for the extra-dimensional part $\psi_{gs}(z)$: 
\begin{equation}
    \label{gs-Slike-equation}
    \left[-\partial_{z}^{2} + V_{gs}(z)\right] \psi_{gs}(z) = m^2 \psi_{gs}(z) ,
\end{equation}
where $m^2=\omega^2-p^2$ is the mass of the KK mode of the scalar perturbation. 
In Sec.~\ref{Quasinormal modes}, we will demonstrate that there is no zero mode for the scalar perturbation, 
thereby confirming the absence of a fifth force on the brane~\cite{Liu:2017gcn}.
Regarding the vector perturbation of the brane, only the zero mode exists. 
However, this mode is not localized on the brane when the four-dimensional Planck mass is finite~\cite{Giovannini:2001fh}.

%%%%%%%%%%%%%%%%%%%%%%%%%%%%%%%%%%%%%%%%%%%%%%%%%%%%%%%%%%%%%%%%%%%%%%%%%%%%%%%%%%%%%%%%%%%%%%%%%%%%%%%%%%%%%%%%%%%%%%%%%%%%%
\section{Test matter fields} \label{matter fields}

For the test matter fields in the bulk, we can neglect their reaction on the background. 
In this section, we will focus on the bulk scalar field and bulk vector field. 
%%%%%%%%%%%%%%%%%%%%%%%%%%%%%%%%%%%%%%%%%%%%%%%%%%%%%%%%%%%%%%%%%%%%%%%%%%%%%%%%%%%%%%%%%%%%%%%%
\subsection{Spin-0 scalar field} \label{Spin-0 scalar fields}

First, we consider the action of a massless scalar field $X_{s}(x^{\mu},z)$: 
\begin{equation}
    \label{Spin0-action}
    S_{0}=-\frac{1}{2} \int dx^4dz \sqrt{-g} \, g^{MN}\partial_{M}X_{s}\partial_{N}X_{s}.  
\end{equation}
The corresponding equation of motion of the scalar field $X_{s}(x^{\mu},z)$ is 
\begin{equation}
    \label{Spin0-EOM}
    \frac{1}{\sqrt{-g}}\partial_{M}\left(\sqrt{-g}\,g^{MN}\partial_{N}X_{s}\right) =0 .
\end{equation}
Using the background metric~\eqref{line-element-z}, we can obtain 
\begin{equation}
    \label{Spin0-EOM-Simp}
    \Box^{(4)}X_{s}+ \partial_{z}^{2} X_{s} + 3\partial_{z}A \partial_{z} X_{s} =0 . 
\end{equation}
Then, one makes the decomposition $X_{s}(x^{\mu},z)= c_{s} e^{-\frac{3}{2}A(z)}e^{-ip_{j}x^{j}} \Psi_{s}(t,z) $, 
where $c_{s}$ is a constant. 
Then, substituting the decomposition into Eq.~\eqref{Spin0-EOM-Simp}, 
we obtain a wave equation for $\Psi_{s}(t,z)$:
\begin{equation}
    \label{Spin0-EOM-wave}
    \left[-\partial_{t}^{2} + \partial_{z}^{2} + V_{s}(z) - p^2\right] \Psi_{s}(t,z)=0 , 
\end{equation}
where
\begin{equation}
    \label{Spin0-EOM-potential}
    V_{s}(z)=\frac{3}{2}\partial_{z}^{2}A + \frac{9}{4}\left(\partial_{z}A\right)^2
\end{equation}
is the effective potential for the scalar field $\Psi_{s}(t,z)$. 
Furthermore, by assuming the decomposition $\Psi_{s}(t,z) = e^{-i\omega t}\psi_{s}(z)$,  
we can obtain a Schrödinger-like equation: 
\begin{equation}
    \label{Spin0-EOM-Slike}
    \left[-\partial_{z}^{2} + V_{s}(z)\right] \psi_{s}(z) = m^2 \psi_{s}(z) ,
\end{equation}
where $m^2=\omega^2-p^2$ is the mass of the KK excitation of the bulk scalar field. 
The zero mode can be expressed as $\psi_{s}^{(0)}(z) \propto e^{\frac{3}{2}A(z)}$. 
Thus, it is easy to show that the scalar zero mode can be localized on the brane for $n>0$. 

%%%%%%%%%%%%%%%%%%%%%%%%%%%%%%%%%%%%%%%%%%%%%%%%%%%%%%%%%%%%%%%%%%%%%%%%%%%%%%%%%%%%%%%%%%%%%%
\subsection{Spin-1 vector field}

Now, we discuss a massless spin-1 vector field $A_{M}(x^{\mu},z)$. 
The action reads: 
\begin{equation}
    \label{spin1-action}
    S_{1}= -\frac{1}{4}\int dx^4dz \sqrt{-g}\,g^{MN}g^{LS}F_{ML}F_{NS}, 
\end{equation}
where $F_{MN}=\partial_{M}A_{N}-\partial_{N}A_{M}$ is the field strength tensor. 
By choosing the gauge $A_{5}=0$, the action~\eqref{spin1-action} can be reduced to
\begin{equation}
    \label{spin1-action-re}
    S_{1}= -\frac{1}{4}\int dx^4dz \sqrt{-g}\left( g^{\mu\nu}g^{\alpha\beta}F_{\mu\alpha}F_{\nu\beta} + 2e^{-A}g^{\mu\nu}\partial_{z}A_{\mu} \partial_{z}A_{\nu} \right).
\end{equation}
Then, we assume the bulk vector field can be decomposed as $A_{\mu}(x^{\alpha},z)=a_{\mu}(x^{\alpha})e^{-\frac{A}{2}}\psi_{v}(z)$. 
By repeating the above steps in Sec.~\ref{Spin-0 scalar fields}, the Schrödinger-like equation satisfied by 
the extra dimensional part $\psi_{v}(z)$ of the vector field can be obtained as 
\begin{equation}
    \label{spin1-Slike-equation}
    \left[-\partial_{z}^{2} + V_{v}(z)\right] \psi_{v}(z) = m^2 \psi_{v}(z) ,
\end{equation}
where
\begin{equation}
    \label{spin1-potential}
    V_{v}(z)=\frac{1}{2}\partial_{z}^{2}A + \frac{1}{4}\left(\partial_{z}A\right)^2 
\end{equation}
is the effective potential of the vector KK mode for the bulk vector field. The zero mode satisfies $\psi_{v}^{(0)}(z) \propto e^{\frac{1}{2}A(z)}$, 
which can also be localized on the brane.  

Combining the results from the above discussions on the perturbations of the thick brane and bulk matter fields, 
we can ultimately obtain a wave equation and a Schrödinger-like equation. 
In particular, the effective potential of the massless bulk scalar field is the same as that of the tensor perturbation of the brane. 
In the next section, we will investigate the QNMs of these fields, focusing on the solution of the Schrödinger-like equations and the numerical evolution of the fields. 

%%%%%%%%%%%%%%%%%%%%%%%%%%%%%%%%%%%%%%%%%%%%%%%%%%%%%%%%%%%%%%%%%%%%%%%%%%%%%%%%%%%%%%%%%%%%%%%%%%%%%%%%%%%%%%%%%%%%%%%%%%%%%
\section{Quasinormal modes of the thick brane} \label{Quasinormal modes}

In this section, we investigate the QNMs of the thick brane and bulk matter fields. 
First, we discuss whether the choice of the parameter $n$ affects the characteristic of the effective potential. 
Using the effective potential of the tensor perturbation as an example, 
we find different behaviors of the potential at the boundary depending on the value of $n$. 
Then, we employ three different methods to calculate QNMs and obtain QNMs of various perturbations of the brane and bulk matter fields. 
Finally, we numerically evolve the wave equations to obtain the waveforms that would be observed in actual scenarios.

%%%%%%%%%%%%%%%%%%%%%%%%%%%%%%%%%%%%%%%%%%%%%%%%%%%%%%%%%%%%%%%%%%%%%%%%%%%%%%%%%%%%%%%5
\subsection{Properties of effective potentials } \label{The properties of effective potentials}

Analogous to the QNM problem of a black hole, in the braneworld scenario, we solve a Schrödinger-like equation 
with appropriate boundary conditions. 
Different types of perturbations correspond to different effective potentials, which govern the characteristics of the QNMs. 

Now we start to consider the case of $n=1$. 
Combining Eq.~\eqref{z-y-1} with Eqs.~\eqref{Ay-expression} and~\eqref{phiy}, 
the specific expressions of the warp factor $A(z)$ and bulk scalar field $\phi(z)$ can be obtained as 
\begin{align}
    \label{Az-n1}
    n=1:\quad A(z) &= -\log \left(\cosh(kz)\right) , \\
    \label{phiz-n1}
    \phi(z) &= \sqrt{3} k z .  
\end{align}
The effective potentials of various perturbations and bulk matter fields are  
\begin{align}
    \label{Vgt-n1}
    n=1:\quad V_{gt} &= V_{s} = \frac{3 k^2}{4} \left( 3-\frac{5}{\cosh^2(k z)} \right)  ,\\
    \label{Vgs-n1}
    V_{gs} &= \frac{k^2}{4} \left( 9+\frac{1}{\cosh^2(k z)} \right)  ,\\
    \label{Ve-n1}
    V_{v}  &= \frac{k^2}{4} \left( 1-\frac{3}{\cosh^2(k z)} \right)  .
\end{align}
Similarly, we can also obtain the results for the case of $n=2$: 
\begin{align}
    \label{Az-n2}
    n=2:\quad A(z) &= -\log \left(1+k^2 z^2\right) , \\
    \label{phiz-n2}
    \phi(z) &= \sqrt{6} \, \log \left(k z + \sqrt{1+k^2 z^2}\right) .  
\end{align}
The corresponding effective potentials are 
\begin{align}
    \label{Vgt-n2}
    n=2:\quad V_{gt} &= V_{s} = \frac{3 k^2 \left(-1+4 k^2 z^2\right) }{\left(1+k^2 z^2\right)^2}   ,\\
    \label{Vgs-n2}
    V_{gs} &= \frac{6 k^2}{1+k^2 z^2}   ,\\
    \label{Ve-n2}
    V_{v}  &= \frac{k^2 \left(-1+2 k^2 z^2\right) }{\left(1+k^2 z^2\right)^2}   .
\end{align}
However, when $n$ is a noninteger, 
obtaining analytical expressions for the solutions and effective potentials is not feasible. 
Therefore, only numerical results can be provided. 

\begin{figure*}[htb]
    \begin{center}
    \subfigure[~Tensor perturbation]  {\label{Vgtz-Plot1}
    \includegraphics[width=5.6cm]{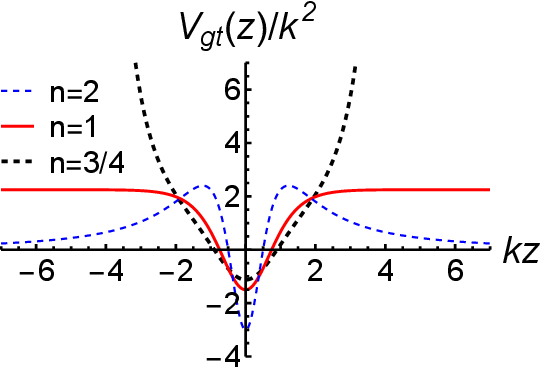}}
    \subfigure[~Scalar perturbation]  {\label{Vgsz-Plot1}
    \includegraphics[width=5.6cm]{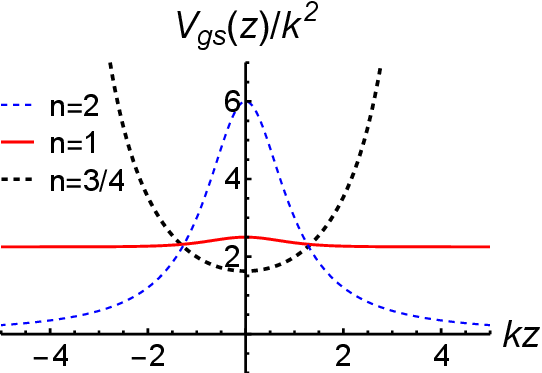}}
    \subfigure[~Bulk vector field]  {\label{Vez-Plot1}
    \includegraphics[width=5.6cm]{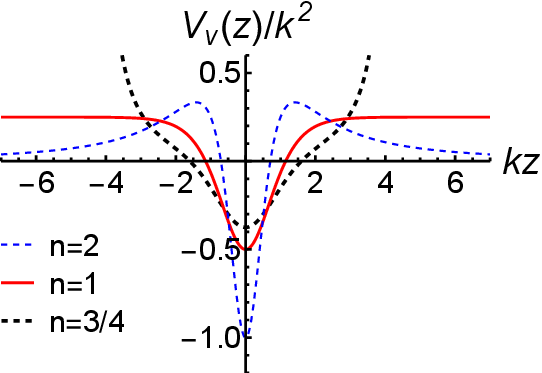}}
    \end{center}
    \caption{The shapes of the effective potentials of the tensor perturbation, the scalar perturbation, and 
        the bulk vector field in the conformal coordinate $z$. }
    \label{Figure-2}
\end{figure*}

We have identified certain universal properties across different types of perturbations and bulk matter fields. 
\begin{itemize}
    \item When $0<n<1$, all of the effective potentials approach infinity at the boundaries, which is similar to the behavior observed in AdS thick branes\cite{Liu:2009uca}. 
    \item When $n=1$, these potentials approach a constant value at the boundaries, resembling the mass gap observed in de Sitter (dS) thick branes\cite{Wang:2002pka}. 
    \item When $n>1$, these potentials approach zero at the boundaries, which is what we commonly refer to as a volcano-like potential, corresponding to Poincaré thick branes\cite{DeWolfe:1999cp,Gremm:1999pj,Csaki:2000fc}. 
\end{itemize}
In particular, we focus on specific values of $n=\tfrac{3}{4}$, $1$, and $2$ for detailed discussions, as shown in Fig.~\ref{Figure-2}. 
For the thick brane with an infinite extra dimension, the choice of the parameter $n$ does not affect the asymptotic behavior of the effective potentials at the boundaries~\cite{Tan:2022vfe,Tan:2023cra}. 

%%%%%%%%%%%%%%%%%%%%%%%%%%%%%%%%%%%%%%%%%%%%%%%%%%%%%%%%%%%%%%%%%%%%%%%%%%%%%%%%%%%%%%%%%%%%
\subsection{The case of \texorpdfstring{$n=1$}{n=1}} \label{the case of n1}

As can be seen from Fig.~\ref{Figure-2} and Eqs.~\eqref{Vgt-n1},~\eqref{Vgs-n1}, and~\eqref{Ve-n1}, when $n=1$, the effective potential is the Pöschl-Teller (PT) potential. 
In the context of black hole QNMs, the PT potential is commonly used as an approximate effective potential for addressing the problem of QNMs. 
Below, we review the bound state solutions of the PT potential (see Refs.~\cite{Gremm:2000dj,Barbosa-Cendejas:2007ucz,D_az_1999}). 

We start from the Schrödinger-like equation: 
\begin{equation}
    \label{PT-equation}
    - \partial_{\bar{z}}^{2} \psi - \frac{\nu(\nu +1)}{\cosh^2 \bar{z}}\psi = E \psi .   
\end{equation}
The necessary condition for the existence of bound state solutions is $\nu >0$. Thus, the eigenvalue $E$ is negative.  
Consequently, the eigenvalue $E$ associated with these solutions is negative.
The general bound state solution is 
\begin{equation}
    \psi(\bar{z}) = \left(\cosh \bar{z}\right)^{\nu+1} \left[C_{1}\cdot  {}_2\!F_1 \left(a,b,c,-\sinh^2 \bar{z}\right) + C_{2}\cdot {}_2\!F_1 \left(a+\frac{1}{2},b+\frac{1}{2},c+1,-\sinh^2 \bar{z}\right) \left(\sinh \bar{z}\right) \right] ,
\end{equation}
where
\begin{equation}
    a = \frac{1}{2}\left(\nu + 1 -\sqrt{-E}\right), \qquad b=\frac{1}{2}\left(\nu + 1 +\sqrt{-E}\right), \qquad c=\frac{1}{2}, 
\end{equation}
and $C_1$ and $C_2$ are constant coefficients. 
Applying the normalization condition, one can obtain the eigenenergy spectrum as follows: 
\begin{equation}
    E_{N}= -\left(\nu - N\right)^2 , \qquad 0\leqslant N < \nu ,~ N=0,1,2,\dots 
\end{equation}
It can be seen that the number of bound states is $[\nu]$, which represents $\nu$ is rounded up. 

For the problem of QNMs, we need to go back to solving the wave equation of the following form with outgoing boundary conditions 
\begin{equation}
    \label{PT-wave-equation}
    \left[\partial_{t}^{2} - \partial_{\bar{z}}^{2} - \frac{\nu(\nu +1)}{\cosh^2 \bar{z}} + p^2\right] \Psi(t,\bar{z})=0 .
\end{equation}
With the ansatz $\Psi(t,\bar{z}) \thicksim  e^{-i\omega t} \psi(\bar{z})$, we can obtain
\begin{equation}
    \label{PT-re-equation}
    \left[- \partial_{\bar{z}}^{2} - \frac{\nu(\nu +1)}{\cosh^2 \bar{z}} \right] \psi(\bar{z})= m^2 \psi(\bar{z}) ,
\end{equation}
where $m^2=\omega^2-p^2$. Correspondingly, the outgoing boundary conditions can be expressed as 
\begin{equation}
    \label{boundary-condition}
    \psi(\bar{z}) \propto \left\{
    \begin{aligned}
        e^{i m \bar{z}} &, \quad \bar{z}\rightarrow \infty, \\
        e^{-i m \bar{z}} &, \quad \bar{z}\rightarrow -\infty. 
    \end{aligned}
    \right.
\end{equation}
Next, we need to perform a coordinate transformation $\hat{z}=\frac{1-\tanh \bar{z}}{2}$, 
then the corresponding boundary conditions are
\begin{equation}
    \label{boundary-condition-re}
    \psi(\hat{z}) \propto \left\{
    \begin{aligned}
        &\hat{z}^{-i m/2} ,  & \hat{z}\rightarrow 0, \\
        &(1-\hat{z})^{-i m/2} ,  & \hat{z}\rightarrow 1.
    \end{aligned}
    \right.
\end{equation}
Then, if we set $\psi(\hat{z})=\hat{z}^{-i m/2} \left(1-\hat{z}\right)^{-i m/2} \xi(\hat{z})$ and substitute it into Eq.~\eqref{PT-re-equation}, 
we can get a standard hypergeometric equation for $\xi(\hat{z})$: 
\begin{equation}
    \label{PT-equation-HS}
    \hat{z}\left(1-\hat{z}\right) \frac{d^2 \xi}{d \hat{z}^2} + \left[\gamma-(\alpha+\beta+1)\hat{z}\right]\frac{d \xi}{d \hat{z}} -\alpha\beta \xi =0, 
\end{equation}
where $\alpha=-im-\nu$, $\beta=-im+\nu+1$, and $\gamma=1-im$. In general, the general solution of Eq.~\eqref{PT-equation-HS} is 
\begin{equation}
    \xi(\hat{z}) = D_{1}\cdot {}_2\!F_1 \left(\alpha,\beta,\gamma,\hat{z}\right) + D_{2}\cdot \hat{z}^{1-\gamma} {}_2\!F_1 \left(\alpha-\gamma+1,\beta-\gamma+1,2-\gamma,\hat{z}\right).
\end{equation}
Thus, the solution of Eq.~\eqref{PT-re-equation} is 
\begin{equation}
    \label{PT-equation-sol-QNMs}
    \psi(\hat{z}) = D_{1}\cdot \left(\hat{z}(1-\hat{z})\right)^{-i m/2} {}_2\!F_1 \left(\alpha,\beta,\gamma,\hat{z}\right)
    + D_{2}\cdot \hat{z}^{i m/2}(1-\hat{z})^{-i m/2} {}_2\!F_1 \left(\alpha-\gamma+1,\beta-\gamma+1,2-\gamma,\hat{z}\right). 
\end{equation}
When $z \rightarrow \infty$ ($\hat{z}\rightarrow 0$), using ${}_2\!F_1 \left(\alpha,\beta,\gamma,0\right)=1$ and $\psi(\hat{z})\thicksim \hat{z}^{-i m/2}$, we can obtain $D_{2} = 0$, so 
\begin{equation}
    \label{PT-equation-sol-QNMs-1}
    \psi(\hat{z}) =D_{1}\cdot \left(\hat{z}(1-\hat{z})\right)^{-i m/2} {}_2\!F_1 \left(\alpha,\beta,\gamma,\hat{z}\right).
\end{equation}
When $z \rightarrow -\infty$ ($\hat{z}\rightarrow 1$), using 
\begin{equation*}
\begin{aligned}
    {}_2\!F_1 \left(\alpha,\beta,\gamma,\hat{z}\right) =& (1-\hat{z})^{i m} \frac{\Gamma (\gamma) \Gamma(\alpha+\beta-\gamma)}{\Gamma(\alpha)\Gamma(\beta)} {}_2\!F_1 \left(\gamma-\alpha,\gamma-\beta,\gamma-\alpha-\beta+1,1-\hat{z}\right) \\
    +& \frac{\Gamma (\gamma) \Gamma(\gamma-\alpha-\beta)}{\Gamma(\gamma-\alpha)\Gamma(\gamma-\beta)} {}_2\!F_1 \left(\alpha,\beta,-\gamma+\alpha+\beta+1,1-\hat{z}\right) 
\end{aligned}
\end{equation*}
and $\psi(\hat{z})\thicksim \left(1-\hat{z}\right)^{-i m/2}$, we require that the first term is zero, that is $\frac{1}{\Gamma(\alpha)}=0$ or $\frac{1}{\Gamma(\beta)}=0$. 
Thus, the QNMs satisfy the following relationship:
\begin{equation}
    m = \pm \sqrt{-\nu(\nu+1)-\frac{1}{4}} - i \frac{2 N +1}{2}, \qquad N=0,1,2,\dots 
\end{equation}
where $N$ is the overtone index. 

\begin{figure*}[htb]
    \begin{center}
    \subfigure[~Tensor perturbation]  {\label{Vpsie0}
    \includegraphics[width=5.6cm]{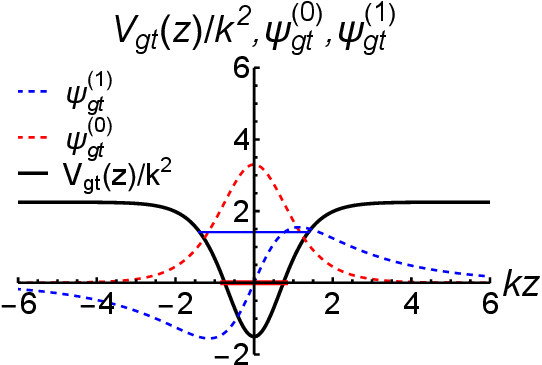}}
    \subfigure[~Bulk vector field]  {\label{Vpsigt01}
    \includegraphics[width=5.6cm]{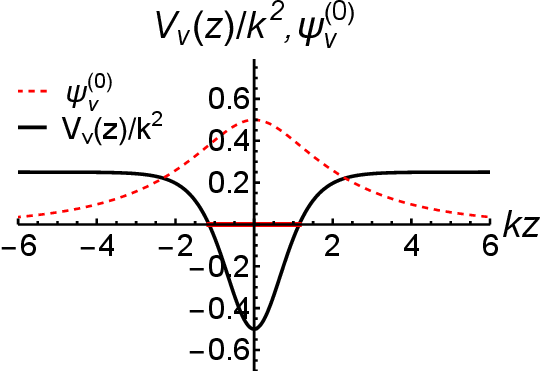}}
    \end{center}
    \caption{The shapes of the effective potential (the black lines)~\eqref{Vgt-n1} and~\eqref{Ve-n1}, the bound state $\psi^{(0)}$ (the red dashed lines), and the bound state $\psi^{(1)}$ (the blue dashed lines)
        for the tensor perturbation and the bulk vector field. The horizontal lines represent the KK mass spectrums. The parameter $n$ is taken as $n=1$.}
    \label{Figure-3}
\end{figure*}

Now, let us return to the discussion on specific issues. 
For the effective potentials~\eqref{Vgt-n1},~\eqref{Vgs-n1}, and~\eqref{Ve-n1},
we can unify them into the form $\frac{V}{k^2} = \alpha - \frac{\beta}{\cosh^2 (kz)} $. 
Then, by performing the coordinate rescaling $\bar{z}=kz$, the corresponding Schrödinger-like equation can be rewritten as 
\begin{equation}
    \label{master-equation}
    \left[ -\partial_{\bar{z}}^{2} - \frac{\beta}{\cosh^2 \bar{z}} \right] \psi(\bar{z}) = \left(\frac{m^2}{k^2} - \alpha\right) \psi(\bar{z}),
\end{equation}
where $\beta=\nu(\nu+1)$. According to $\left[\nu_{gt}\right]=\left[\tfrac{3}{2}\right]=2$, $\left[\nu_{gs}\right]=\left[-\tfrac{1}{2}\right]=0$, and $\left[\nu_{v}\right]=\left[\tfrac{1}{2}\right]=1$, 
we see that the bound states and KK mass spectrums corresponding to various types of effective potentials are listed as follows: 
\begin{itemize}
    \item Tensor perturbation and bulk scalar field (see Fig.~\ref{Vpsie0}): \par
        There are two bound states: 
        \begin{align}
            &m_{0} = 0 , \qquad\qquad \;\;  \psi_{gt}^{(0)}(\bar{z}) \thicksim \left(\cosh(\bar{z})\right)^{-\frac{3}{2}}, \\
            &m_{1} = \sqrt{2}k , \qquad\quad  \psi_{gt}^{(1)}(\bar{z}) \thicksim \frac{\sinh(\bar{z})}{\left(\cosh(\bar{z})\right)^{\frac{3}{2}}} . 
        \end{align}
    \item Scalar perturbation: \par
        There is no bound state. 
    \item Bulk vector field (see Fig.~\ref{Vpsigt01}): \par
        There is only one bound state, i.e., the zero mode: 
        \begin{equation}
            m_{0} = 0 , \qquad\quad \psi_{v}^{(0)}(\bar{z}) \thicksim \left(\cosh(\bar{z})\right)^{-\frac{1}{2}} . 
        \end{equation} 
\end{itemize}
\begin{figure*}[htb]
    \begin{center}
    \subfigure[~Tensor perturbation]  {\label{mgtPTS}
    \includegraphics[width=5.6cm]{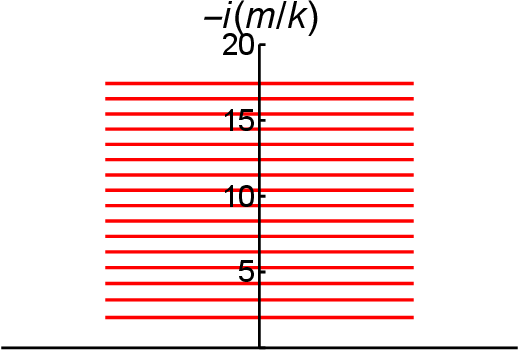}}
    \subfigure[~Scalar perturbation]  {\label{mgsPTS}
    \includegraphics[width=5.6cm]{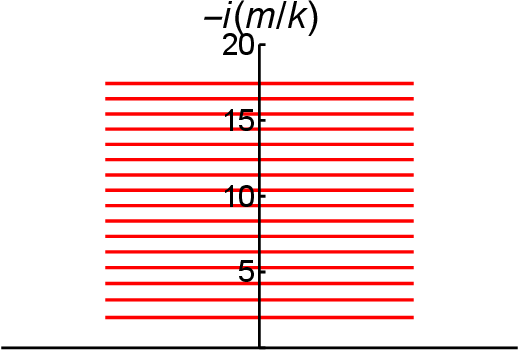}}
    \subfigure[~Bulk vector field]  {\label{mePTS}
    \includegraphics[width=5.6cm]{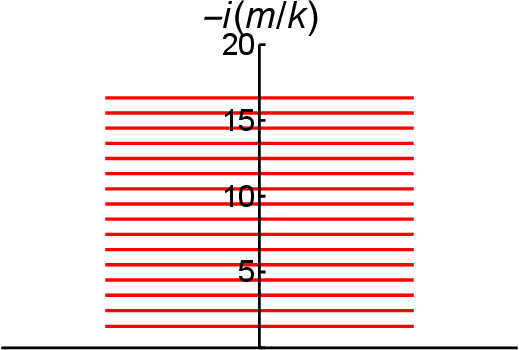}}
    \end{center}
    \caption{The quasinormal spectrum for each type of effective potential for $n=1$. }
    \label{Figure-qnm-n1}
\end{figure*}
And we can calculate the QNMs for each type of effective potential. 
The results are listed as follows: 
\begin{itemize}
    \item Tensor perturbation and bulk scalar field (see Fig.~\ref{mgtPTS}): 
        \begin{equation}
            \label{result-tensor-perturbation}
            \frac{m}{k} = \left\{-2i,-\sqrt{10}i,-3\sqrt{2}i,-2\sqrt{7}i,\cdots,-\sqrt{(N+4)(N+1)}i,\cdots\right\}. 
        \end{equation}
    \item Scalar perturbation (see Fig.~\ref{mgsPTS}): 
        \begin{equation}
            \label{result-scalar-perturbation}
            \frac{m}{k} = \left\{-2i,-\sqrt{10}i,-3\sqrt{2}i,-2\sqrt{7}i,\cdots,-\sqrt{(N+4)(N+1)}i,\cdots\right\}. 
        \end{equation}
    \item Bulk vector field (see Fig.~\ref{mePTS})
        \begin{equation}
            \label{result-bulk-vector}
            \frac{m}{k} = \left\{-\sqrt{2}i,-\sqrt{6}i,-2\sqrt{3}i,-2\sqrt{5}i,\cdots,-\sqrt{(N+2)(N+1)}i,\cdots\right\}. 
        \end{equation}
\end{itemize}

To verify the analytical results~\eqref{result-tensor-perturbation},~\eqref{result-scalar-perturbation}, and~\eqref{result-bulk-vector}, 
we use the continued fraction method~\cite{Leaver:1985ax,Leaver:1990zz} to calculate the QNMs of various perturbations and bulk fields for the case of $n=1$.
Namely, we need to solve the Schrödinger-like equation~\eqref{master-equation} with the outgoing boundary conditions
\begin{equation}
    \label{boundary-condition-re1}
    \psi(\bar{z}) \propto \left\{
    \begin{aligned}
        e^{i \bar{m} \bar{z}} &, \quad \bar{z}\rightarrow \infty, \\
        e^{-i \bar{m} \bar{z}} &, \quad \bar{z}\rightarrow -\infty, 
    \end{aligned}
    \right.
\end{equation}
where $\bar{m}^2 = \frac{m^2}{k^2} - \alpha $. 

First, by performing a coordinate transformation $\tilde{z}=\tanh \bar{z}$, 
Eq.~\eqref{master-equation} and the corresponding boundary conditions~\eqref{boundary-condition-re1} can be rewritten as  
\begin{equation}
    \label{master-equation-re}
    \left\{(1-\tilde{z}^2) \partial_{\tilde{z}}^2-2\tilde{z}\partial_{\tilde{z}} +\left[\frac{\bar{m}^2}{1-\tilde{z}^2} +\beta \right]\right\}\psi(\tilde{z})=0, 
\end{equation}
and
\begin{equation}
    \label{boundary-condition-re2}
    \psi(\hat{z}) \propto \left\{
    \begin{aligned}
        &(1+\tilde{z})^{-i \bar{m}/2} ,  & \tilde{z}\rightarrow -1, \\
        &(1-\tilde{z})^{-i \bar{m}/2} ,  & \tilde{z}\rightarrow 1,  
    \end{aligned}
    \right.
\end{equation}
respectively. Then, following the continued fraction method proposed by Leaver~\cite{Leaver:1985ax}, 
the ansatz satisfying the boundary conditions~\eqref{boundary-condition-re2} for the wave function $\psi(\tilde{z})$ is
\begin{equation}
    \psi(\tilde{z}) = (1+\tilde{z})^{-i \bar{m}/2} (1-\tilde{z})^{-i \bar{m}/2} \sum_{l=0}^{\infty}a_l \tilde{z}^l ,
\end{equation}
where $a_l$ is the expansion coefficient. 
Substituting this into Eq.~\eqref{master-equation-re}, a two-term recurrence relation can be obtained:
\begin{equation}
    \label{recurrence-relation}
    \gamma_l a_{l+2} + \delta_l a_l =0,\quad l=0,1,2,\cdots
\end{equation}
where the recurrence coefficients $\gamma_l$ and $\delta_l$ are functions of $l$ and the parameter $\beta$: 
\begin{align}
    \gamma_l &= 2 + l - (2 + l)^2 , \\
    \delta_l &= (l^2 - \beta + l (1 - 2 i \bar{m}) - \bar{m} (i + \bar{m})). 
\end{align}
In fact, the recurrence relation~\eqref{recurrence-relation} can be written into the following matrix form: 
\begin{equation}
    \label{matrix-equation}
    \begin{pmatrix}
        \delta_0 & 0 & \gamma_0 & & & \\
         & \delta_1 & 0 & \gamma_1 & & \\
         &  & \delta_2 & 0 & \gamma_2 & \\
         &  &  & \ddots & \ddots & \ddots
    \end{pmatrix}
    \begin{pmatrix}
        a_0 \\
        a_1 \\
        a_2 \\
        \vdots 
    \end{pmatrix}
    =0. 
\end{equation}
It can be seen that this matrix equation~\eqref{matrix-equation} is essentially an algebraic equation for the QNMs. 
Solving this linear system yields the QNMs, and this system has nontrivial solution if and only
if the determinant of the coefficient matrix is zero.

Generally, solving the eigenvalue problem for this matrix equation involves dealing with an infinite-dimensional secular equation. 
To make the problem solvable, we can truncate it at a sufficiently large dimension, and the larger the dimension of the matrix, the more accurate the results.
Fortunately, since this coefficient matrix is an upper triangular matrix, we can directly obtain: 
\begin{equation}
    \delta_0=0\;,\; \delta_1=0\;,\; \delta_2=0\;,\; \cdots\;,\; \delta_l=0 \;,\; \cdots 
\end{equation}
namely, 
\begin{equation}
    \bar{m} = \frac{1}{2} \left[-i(1+2l) \pm \sqrt{-1 - 4 \beta}\right],\quad l=0,1,2,\cdots. 
\end{equation}
Finally, through the relation $\bar{m}^2 = \frac{m^2}{k^2} - \alpha $, 
we find that the results are completely consistent with those obtained from the analytical solution.

From these results, it can be observed that the QNMs are all pure imaginary, 
implying that these modes decay purely without oscillation. 
This behavior is similar to the results of gravitational perturbations in the background of odd-dimensional dS spacetime~\cite{Natario:2004jd}. 

%%%%%%%%%%%%%%%%%%%%%%%%%%%%%%%%%%%%%%%%%%%%%%%%%%%%%%%%%%%%%%%%%%%%%%%%%%%%%%%%%%%%%%%%%%%%%%%%%%%%%%%
\subsection{The case of \texorpdfstring{$n>1$}{n>1}}

For the case of $n=2$, there are no analytical solutions available. 
Similar to solving the QNMs of a black hole, various semi-analytical methods can be employed, 
including the Wentzel-Kramers-Brillouin (WKB) method~\cite{Schutz:1985km}, the asymptotic iteration method (AIM)~\cite{Tan:2022vfe,Cho:2011sf}, the shooting method~\cite{Pani:2013pma}, 
and so on, to solve the QNMs of the thick brane. 

In contrast to the effective potential of a pure potential barrier for a black hole, there is a potential well 
for the tensor perturbation and the bulk vector field for the thick brane considered in this paper. 
However, we can use the method of supersymmetric quantum mechanics to calculate the dual potential corresponding to the effective potential. 
Reference~\cite{Cooper:1994eh} pointed out that the effective potential and the dual potential share the same QNM spectrum of excitation states. 

\begin{figure*}[htb]
    \begin{center}
    \subfigure[~Tensor perturbation]  {\label{AllVgtz2P}
    \includegraphics[width=5.6cm]{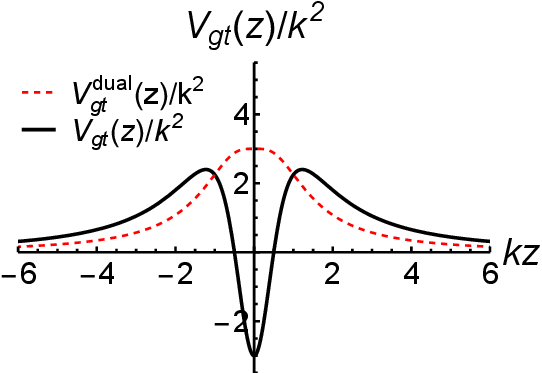}}
    \subfigure[~Bulk vector field]  {\label{AllVez2P}
    \includegraphics[width=5.6cm]{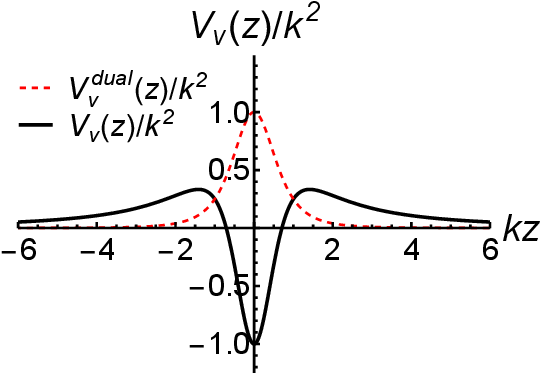}}
    \end{center}
    \caption{The shapes of the effective potentials (the black lines) and the dual potentials (the red dashed lines) for $n=2$.}
    \label{Figure-4}
\end{figure*}

Now we consider the supersymmetric form of the Schrödinger-like equation~\eqref{gt-Slike-equation}:
\begin{equation}
    \label{supersymmetric-equation}
    \mathcal{O}\mathcal{O}^{\dagger} \psi_{gt}(z) = \left[-\partial_{z}^{2} + V_{gt}(z)\right] \psi_{gt}(z) = m^2 \psi_{gt}(z), 
\end{equation}
where 
\begin{eqnarray}
    \mathcal{O} &=&\partial_{z}+\frac{3}{2}\partial_{z}A ,\\
    \mathcal{O}^{\dagger} &=&-\partial_{z}+\frac{3}{2}\partial_{z}A , \\
    \label{Vgt2}
    V_{gt}(z) &=& \frac{3}{2}\partial_{z}^{2}A + \frac{9}{4}\left(\partial_{z}A\right)^2 = \frac{3 k^2 \left(-1+4 k^2 z^2\right) }{\left(1+k^2 z^2\right)^2}. 
\end{eqnarray}
According to supersymmetric quantum mechanics, the above equation has a dual equation:
\begin{equation}
    \label{supersymmetric-equation-dual}
    \mathcal{O}^{\dagger}\mathcal{O} \bar{\psi}_{gt}(z) = \left[-\partial_{z}^{2} + V_{gt}^{\text{dual}}(z)\right] \bar{\psi}_{gt}(z) = m^2 \bar{\psi}_{gt}(z) ,
\end{equation}
where 
\begin{equation}
    \label{gt-effective-potential-dual}
    V_{gt}^{\text{dual}}(z) = -\frac{3}{2}\partial_{z}^{2}A + \frac{9}{4}\left(\partial_{z}A\right)^2 = \frac{3 k^2\left(1+2k^2 z^2\right)}{\left(1+k^2 z^2\right)^2}
\end{equation}
is the dual potential corresponding to the effective potential~\eqref{Vgt2} (see Fig.~\ref{AllVgtz2P}). 
Similarly, one can also obtain the dual potential of the effective potential~\eqref{Ve-n2} for the bulk vector field as (see Fig.~\ref{AllVez2P})
\begin{equation}
    \label{e-effective-potential-dual}
    V_{v}^{\text{dual}}(z) = -\frac{1}{2}\partial_{z}^{2}A + \frac{1}{4}\left(\partial_{z}A\right)^2 = \frac{k^2}{\left(1+k^2 z^2\right)^2}.
\end{equation}

We plot the effective potentials and the dual ones of the tensor perturbation and the bulk vector field in Fig.~\ref{Figure-4}. 
From this figure, we can see that the shapes of these dual potentials are similar to the effective potentials of black holes.  
Thus, we can use the AIM, the shooting method, and the WKB method 
to calculate the quasinormal frequencies (QNFs) for the potentials~\eqref{gt-effective-potential-dual},~\eqref{Vgs-n2}, and~\eqref{e-effective-potential-dual}. 
The results are listed in Tab.~\ref{Table-1}. 
\begin{center}
    \begin{table}[!htb]
    \renewcommand\arraystretch{0.8}
    \begin{tabular}{|c|c|c|c|c|}
    \hline
    $V(z)$   &~$N$~     & AIM    &~Shooting method~  & WKB method \\
    \hline
        & &~$\text{Re}(m/k) $~~~~$\text{Im}(m/k) $~&~$\text{Re}(m/k) $~~~~$\text{Im}(m/k) $~&~$\text{Re}(m/k) $~~~~$\text{Im}(m/k) $~ \\
    \hline
                                    & 1&    1.73769~~~~-0.30514&1.73769~~~~-0.30514 & $\setminus$ \\
        $V_{gt}^{\text{dual}}(z)$   & 2&    1.72028~~~~-1.03960&1.72030~~~~-1.03960 & $\setminus$ \\
                                    & 3&    1.52494~~~~-2.03052& $\setminus$ & $\setminus$ \\
    \hline
                                    & 0&    2.35494~~~~-0.47773&2.35495~~~~-0.47773 & 2.35260 ~~~~ -0.48132\\
        $V_{gs}(z)$                 & 1&    2.15924~~~~-1.46306&2.15920~~~~-1.46308 & 2.15471 ~~~~ -1.48249 \\
                                    & 2&    1.76893~~~~-2.56023& $\setminus$ & 1.77025 ~~~~ -2.60541 \\
    \hline
                                    & 1&    0.68136~~~~-0.61775&0.68136~~~~-0.61779& 0.66053 ~~~~-0.55112 \\
        $V_{v}^{\text{dual}}(z)$    & 2&    0.46717~~~~-2.17646&  $\setminus$ & 0.42751 ~~~~-2.15338 \\
                                    & 3&    0.41088~~~~-3.79665& $\setminus$ & $\setminus$ \\
    \hline

    \end{tabular}\\
    \caption{Low overtone QNFs calculated with the AIM, the shooting method, and the WKB method.}
    \label{Table-1}
    \end{table}
\end{center}

In Tab.~\ref{Table-1}, the symbol ``$\setminus$'' indicates that the method is unable to achieve the required computational accuracy 
and thus produces erroneous values, which are excluded.  
From this table, we can see that the results of the first two methods are consistent. 
It is important to note that unlike the case of the black hole, the effective potentials of the tensor perturbation and the bulk vector field in the braneworld support bound state zero modes. 
Therefore, in addition to the absence of a zero mode in the scalar perturbation, we denote the first quasinormal frequency as the overtone $N=1$. 

\begin{figure*}[htb]
    \begin{center}
    \subfigure[~Tensor perturbation]  {\label{nonVgtz2P}
    \includegraphics[width=5.6cm]{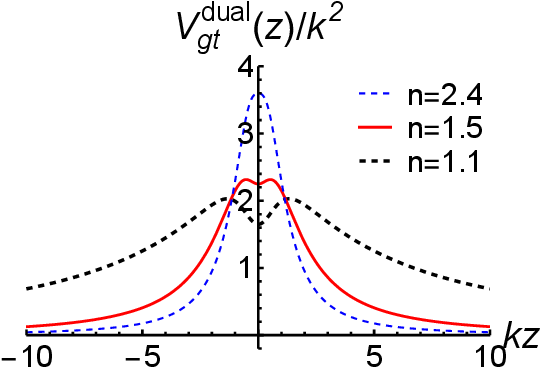}}
    \subfigure[~Scalar perturbation]  {\label{nonVgsz2P}
    \includegraphics[width=5.6cm]{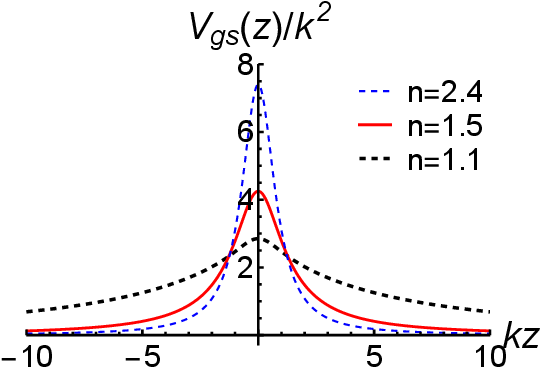}}
    \subfigure[~Bulk vector field]  {\label{nonVez2P}
    \includegraphics[width=5.6cm]{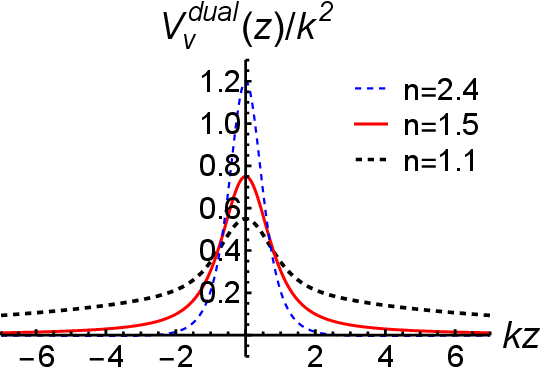}}
    \end{center}
    \caption{The behavior of the dual potential or the effective potential with different values of $n$. }
    \label{Figure-5}
\end{figure*}

For the case of $n>1$ and $n$ is a noninteger, we are unable to obtain the inverse function $y=y(z)$ analytically for the function~\eqref{z-y-n}. 
Therefore, one can only use numerical methods to fit the approximate expressions of the effective potentials in the $z$-coordinate. 
First, we discuss how the effective potentials or the dual ones change with the parameter $n$.  
From Fig.~\ref{Figure-5}, we can see that as $n$ approaches $1$, the dual potential of the tensor perturbation in Fig.~\ref{nonVgtz2P} exhibits a small quasi-potential well around $z=0$. 
This indicates the presence of a potentially ``long-lived'' mode in this situation~\cite{Tan:2023cra}.

\begin{figure*}[htb]
    \begin{center}
    \subfigure[~Real parts]  {\label{VPnRe}
    \includegraphics[width=5.6cm]{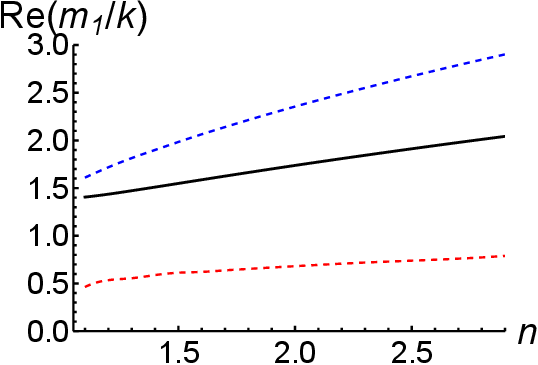}}
    \subfigure[~Imaginary parts]  {\label{VPnIm}
    \includegraphics[width=5.6cm]{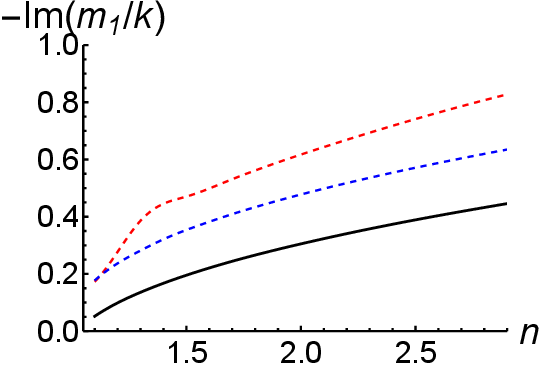}}
    \end{center}
    \caption{The relations between the real/imaginary parts of the first QNFs and the parameter $n$ 
        for the tensor perturbation (the black lines), bulk vector field (the red dashed lines), and scalar perturbation (the blue dashed lines). }
    \label{Figure-6}
\end{figure*}

Then, we plot the variation of the first QNFs with respect to the parameter $n$ in Fig.~\ref{Figure-6}. 
It can be observed that both the absolute values of the real and imaginary parts of the first QNFs increase with $n$. 

From Fig.~\ref{VPnIm}, it can be seen that as $n$ approaches $1$, 
the imaginary part of the first QNF for gravitational tensor perturbation will gradually approach zero.  
The magnitude of the imaginary part of a QNF is related to the lifetime of the QNM. 
Next, we will discuss only the first QNF. 
This is because, in typical experimental observations, 
gravitational wave signals are usually dominated by the first QNM, 
while the second and subsequent QNMs decay faster than the first one and are obscured. 
We can clearly see this phenomenon through the evolution of an initial wave packet in bulk spacetime. 
Now, in the following, we will specifically discuss this phenomenon for $n=1.1$.

\begin{figure*}[htb]
    \begin{center}
    \subfigure[~$\psi$]  {\label{aoddpofilez3}
    \includegraphics[width=5.6cm]{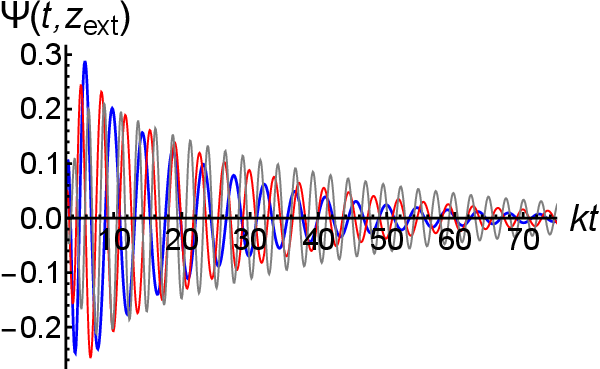}}
    \subfigure[~$|\psi|$]  {\label{alogoddpofilez3}
    \includegraphics[width=5.6cm]{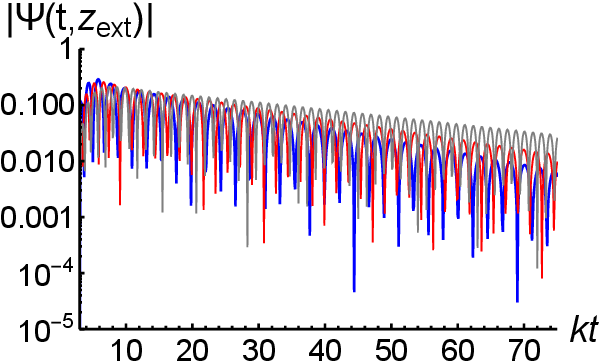}}
    \end{center}
    \caption{Time evolution of the odd initial wave packet at the extraction point $k z_{ext}=3$ 
        for the cases of $p=0$ (the blue lines), $p=1$ (the red lines), and $p=2$ (the gray lines) for $n=1.1$. }
    \label{Figure-7}
\end{figure*}

Using the light-cone coordinates $du=dt-dz$ and $dv=dt+dz$, we can rewrite 
Eq.~\eqref{gt-Phi-equation} as~\cite{Gundlach:1993tp}
\begin{equation}
    \label{uv-equation}
    \left(4 \frac{\partial^{2}}{\partial_{u}\partial_{v}} + V_{gt}(u,v) + p^2 \right)\Psi(u,v)=0. 
\end{equation}
Due to the $z_2$ symmetry of the effective potential, all QNMs are either odd or even. 
When the zero mode exists, the first QNM is odd. 
Therefore, we choose the initial incident wave packet to be an odd function with the following form: 
\begin{align}
    &\Psi_{\text{in}}(u,0)=\sin\left(k u\right) e^{-\frac{k^2 u^2}{2}}, \\
    &\Psi_{\text{in}}(0,v)=\sin\left(k v\right) e^{-\frac{k^2 v^2}{2}}.
\end{align}
With the above initial functions, we plot the evolution of the waveform in Fig.~\ref{Figure-7}. 
\begin{center}
    \begin{table}[!htb]
    \renewcommand\arraystretch{0.8}
    \begin{tabular}{|c|c|c|}
    \hline
    ~~$p/k$~~  & AIM & Time-domain evolution method \\
    \hline
        &~$\text{Re}(\omega_{1}/k) $~~~~$\text{Im}(\omega_{1}/k) $~&~$\text{Re}(\omega_{1}/k) $~~~~$\text{Im}(\omega_{1}/k) $~ \\
    \hline
    0   &   1.40578~~~~-0.0520124&1.40910~~~~-0.0524394 \\
    \hline
    1.0 &   1.72491~~~~-0.0423895&1.72547~~~~-0.0420582 \\
    \hline
    2.0 &   2.44426~~~~-0.0299142&2.44525~~~~-0.0293063 \\
    \hline
    \end{tabular}\\
    \caption{The first frequency $\omega_{1}/k$ using the AIM and time-domain evolution method for $n=1.1$.}
    \label{Table-2}
    \end{table}
\end{center}

We list the numerical results of the ``long-lived'' mode for the gravitational perturbation in Tab.~\ref{Table-2}. 
The frequencies of the QNMs are determined using the time-domain evolution method, where the waveform data is fitted 
with an appropriate wave function: $e^{\text{Im}(\frac{\omega_{1}}{k})t}\sin\left(\text{Re}(\frac{\omega_{1}}{k})t+c_0\right)$. 
For comparison, the results obtained from the AIM are listed in the adjacent column, where $\omega_{1}=\sqrt{m_{1}^{2}+p^2}$. 
By combining Tab.~\ref{Table-2} and Fig.~\ref{Figure-7}, it can be observed that as the three-dimensional wavenumber $p$ increases, 
the imaginary part of the frequency decreases while the real part increases, which is consistent with $\omega_{n}=\sqrt{m^{2}_{n}+p^2}$. 
Furthermore, through numerical calculations, it is found that when $n=1.1$, the frequency of the first QNM is $\omega/k = 1.4091-0.0524394i$ for $p=0$. 
Compared to the result for $n=2$ and $p=0$, where $\omega/k = 1.73769-0.30514i$ as shown in Tab.~\ref{Table-1}, this mode is ``long-lived'', which is consistent with our predicted result.
Combining the above results, we can speculate that as $n$ approaches $1$, the lifetime of this mode will become longer and longer. 
Moreover, it transitions from a quasi-bound mode to a bound state, indicating that for $n=1$, 
there exists a second bound state of the tensor perturbation besides the zero mode as discussed in Sec.~\ref{the case of n1}.

Finally, let us discuss the lifetime of the first QNM. 
First, we see that the characteristic energy scale $k$ within the bulk spacetime is determined by 
the upper limit established through four-dimensional Newtonian potential experiments~\cite{Tu:2007zz,Yang:2012zzb,Tan:2016vwu,Tan:2020vpf,Lee:2020zjt,Du:2022veu,Sui:2020atb}. 
Currently, the upper bound of this measurement corresponds to the submillimeter length scale, 
which is associated with an energy scale of approximately $k \sim 10^{-3}~\text{eV}$. 
For $p=0$, if we assume $k=10^{-3}~\text{eV}$, the half-life of this first QNM is $8.724\times 10^{-12}$ seconds, which is larger than that in the RS-\uppercase\expandafter{\romannumeral2} model, $4\times 10^{-13}$ seconds~\cite{Seahra:2005wk}. 
As this upper limit decreases, i.e., the extra dimension characteristic length scale $l<0.1~\text{mm}$ ($k>10^{-3}~\text{eV}$), the lifetime of the QNM becomes shorter. 
On the other hand, for $p \neq 0$, if we assume $\frac{p}{m} \gg 1$ and $\text{Im}(\omega/k) \sim -10^{-10}$, 
the corresponding half-life is on the order of milliseconds, and the oscillation frequency is also on the order of $10^{8}~\text{Hz}$.
This phenomenon is expected to be detected in the high-frequency components of stochastic gravitational wave background signals~\cite{Li:2017jcz,Zhao:2021zlr}. 
References~\cite{Seahra:2005iq,Seahra:2005wk} indicated that the optimal locations for searching for such high-frequency gravitational wave signals are high-energy
regions associated with the early universe. 

%%%%%%%%%%%%%%%%%%%%%%%%%%%%%%%%%%%%%%%%%%%%%%%%%%%%%%%%%%%%%%%%%%%%%%%%%%%%%%%%%%%%%%%%%%%%%%%%%%%%%%%%%%%%%%%%%%%%%%%%%%%%%
\section{Conclusion} \label{conclusion}

In this paper, we investigated the QNMs of the thick brane with a finite extra dimension. 
Specifically, we separately considered the transverse-traceless tensor perturbation and scalar perturbation of the background, 
as well as a test bulk scalar field (spin-0) and a bulk vector field (spin-1) in the thick brane. 
From the results shown in Fig. \ref{Figure-2}, we found that the parameter $n$ in the warp factor determines the shapes 
of the effective potentials, which can be mainly divided into three categories: 
volcano-like potential, PT potential, and harmonic oscillator potential corresponding to $n>1$, $n=1$, and $0<n<1$, respectively. 
We also encountered similar potentials in different types of thick branes with an infinite extra dimension: 
Poincaré thick brane (volcano-like potential), dS thick brane (PT potential), and AdS thick brane (harmonic oscillator potential). 
Through analytical, semi-analytical, and numerical methods, 
we studied the QNMs for different types of perturbations and bulk fields.

For the csse of $0<n<1$, there are no QNMs for any type of perturbation or bulk field; instead, there is a series of normal modes. 
For $n=1$, the tensor perturbation contains two normal modes, namely, the zero mode and the first excited state, 
as well as a series of separated pure imaginary QNMs. The scalar perturbation, on the other hand, 
only contains a series of separated pure imaginary QNMs. 
Similarly, for the test matter fields, there is both a zero mode and a series of separated pure imaginary QNMs. 
These pure imaginary QNMs also exist in black hole physics~\cite{Cook:2016fge}. 
For $n>1$, as shown in Tab.~\ref{Table-1}, the results obtained by various methods are consistent with each other. 
We found that as the parameter $n$ approaches the critical value $1$, the absolute values of
the real and imaginary parts of the QNMs decrease accordingly. For the tensor perturbation, there exists a ``long-lived'' mode.  
By performing the time-domain evolution of the initial wave packet for the analogous wave equation, 
we found that the change in the three-dimensional wavenumber $p$ affects the lifetime of this mode. 
As $p$ increases, the imaginary part of QNF decreases, and consequently, the lifetime of this QNM also increases. 

Similar to the results obtained in the RS-\uppercase\expandafter{\romannumeral2} thin brane, the QNMs of the thick brane are short-lived. 
It seems that current laser interferometer gravitational wave detectors have no way to detect these QNMs. 
But for the case of nonvanishing spatial three-momentum $p^2 \neq 0$ or $n$ close to $1$, the QNMs will exhibit a slowly oscillating decay tail which can be seen in Fig.~\ref{Figure-7}. 
According to Ref.~\cite{Konoplya:2023fmh}, these tails may have an impact on the stochastic gravitational wave detection experiments 
such as the Pulsar Timing Array experiment~\cite{NANOGrav:2023gor,Koyama:2004cf,Caprini:2018mtu}.
Due to the unique properties of extra dimensions, this study may provide a new way to detect the existence of extra dimensions. 

In the future, we can also study the ``ring" characteristics exhibited by other braneworld models, 
and consider the QNMs of other fields in the braneworld background, thereby enriching the research on braneworld theories. 

%%%%%%%%%%%%%%%%%%%%%%%%%%%%%%%%%%%%%%%%%%%%%%%%%%%%%%%%%%%%%%%%%%%%%%%%%%%%%%%%%%%%%%%%%%%%%%%%%%%%%%%%%%%%%%%
\section*{Acknowledgments}
We would like to thank Yu-Peng Zhang, Chun-Chun Zhu and Wen-Yi Zhou for very useful discussions. 
This work was supported by 
the National Key Research and Development Program of China (Grant No. 2021YFC2203003), 
the National Natural Science Foundation of China (Grants No. 12475056, No. 12205129, No. 12247101, and No. 12347111), 
the China Postdoctoral Science Foundation (Grant No. 2023M741148), 
the 111 Project (Grant No. B20063), 
the Postdoctoral Fellowship Program of CPSF (Grant No. GZC20240458), 
the Major Science and Technology Projects of Gansu Province, 
and Lanzhou City's scientific research funding subsidy to Lanzhou University.
\par


\begin{thebibliography}{99}

%%%%%%%%%%%%%%%%%%%%%%%%%%%%%%%%%%%%%%%%%%%%%%%%5
% GWs

\bibitem{LIGOScientific:2016aoc}
B.~P.~Abbott \textit{et al.} [LIGO Scientific and Virgo],
\emph{{Observation of Gravitational Waves from a Binary Black Hole Merger}},
{\emph{Phys. Rev. Lett.}  {\bfseries 116},  061102 (2016)},
[{{\ttfamily arXiv:1602.03837}}].

%%%%%%%%%%%%%%%%%%%%%%%%%%%%%%%%%%%%%%%%%%%%%%%%%%%%%%%%%%%%%%%%%%%%%%%%%%%%%%%%%%%%%
% the research of BH QNMs 

\bibitem{Berti:2009kk}
E.~Berti, V.~Cardoso, and A.~O.~Starinets,
\emph{{Quasinormal modes of black holes and black branes}},
{\emph{Class. Quant. Grav.} {\bfseries 26}, 163001 (2009)},
[{{\ttfamily arXiv:0905.2975}}].

\bibitem{Cardoso:2016rao}
V.~Cardoso, E.~Franzin, and P.~Pani,
\emph{{Is the gravitational-wave ringdown a probe of the event horizon?}}
{\emph{Phys. Rev. Lett.} {\bfseries 116}, 171101  (2016)},
[{{\ttfamily  arXiv:1602.07309}}].

\bibitem{Kokkotas:1999bd}
K.~D.~Kokkotas and B.~G.~Schmidt,
\emph{{Quasinormal modes of stars and black holes}},
{\emph{Living Rev. Rel.} {\bfseries 2}, 2 (1999)},
[{{\ttfamily  arXiv:gr-qc/9909058}}].

\bibitem{Konoplya:2011qq}
R.~A.~Konoplya and A.~Zhidenko,
\emph{{Quasinormal modes of black holes: From astrophysics to string theory}},
{\emph{Rev. Mod. Phys.} {\bfseries 83}, 793 (2011)},
[{{\ttfamily  arXiv:1102.4014}}].

\bibitem{Jusufi:2020odz}
K.~Jusufi, M.~Azreg-A\"\i{}nou, M.~Jamil, S.-W.~Wei, Q.~Wu, and A.-Z.~Wang,
\emph{{Quasinormal modes, quasiperiodic oscillations, and the shadow of rotating regular black holes in nonminimally coupled Einstein-Yang-Mills theory}},
{\emph{Phys. Rev. D} {\bfseries 103},  024013 (2021)},
[{{\ttfamily  arXiv:2008.08450}}].

\bibitem{Nollert:1999ji}
H.~P.~Nollert,
\emph{{TOPICAL REVIEW: Quasinormal modes: the characteristic `sound' of black holes and neutron stars}},
{\emph{Class. Quant. Grav.} {\bfseries 16}, R159 (1999)}.

\bibitem{Cheung:2021bol}
M.~H.~Y.~Cheung, K.~Destounis, R.~P.~Macedo, E.~Berti, and V.~Cardoso,
\emph{{Destabilizing the Fundamental Mode of Black Holes: The Elephant and the Flea}},
{\emph{Phys. Rev. Lett.} {\bfseries 128}, 111103 (2022)},
[{{\ttfamily  arXiv:2111.05415}}].

%%%%%%%%%%%%%%%%%%%%%%%%%%%%%%%%%%%%%%%%%%%%%%%%%%%%%%%%%%%%%%%%%%%%%%%%%%%%%%%%%%%%%
% extra dimension

\bibitem{Chakraborty:2017qve}
S.~Chakraborty, K.~Chakravarti, S.~Bose, and S.~SenGupta,
\emph{{Signatures of extra dimensions in gravitational waves from black hole quasinormal modes}},
{\emph{Phys. Rev. D} {\bfseries 97}, 104053 (2018)},
[{{\ttfamily  arXiv:1710.05188}}].

\bibitem{Prasobh:2014zea}
C.~B.~Prasobh and V.~C.~Kuriakose,
\emph{{Quasinormal Modes of Lovelock Black Holes}},
{\emph{Eur. Phys. J. C} {\bfseries 74}, 3136 (2014)},
[{{\ttfamily  arXiv:1405.5334}}].

\bibitem{Dey:2020pth}
R.~Dey, S.~Biswas, and S.~Chakraborty,
\emph{{Ergoregion instability and echoes for braneworld black holes: Scalar, electromagnetic, and gravitational perturbations}},
{\emph{Phys. Rev. D} {\bfseries 103}, 084019 (2021)},
[{{\ttfamily  arXiv:2010.07966}}].

\bibitem{Hashemi:2019jlt}
S.~S.~Hashemi, M.~Kord Zangeneh, and M.~Faizal,
\emph{{Charged scalar quasinormal modes for higher-dimensional Born\textendash{}Infeld dilatonic black holes with Lifshitz scaling}},
{\emph{Eur. Phys. J. C} {\bfseries 80}, 111 (2020)},
[{{\ttfamily  arXiv:hep-th/1901.11367}}].

\bibitem{Chen:2016qii}
C.-H.~Chen, H.~T.~Cho, A.~S.~Cornell, and G.~Harmsen,
\emph{{Spin- 3/2 fields in $D$-dimensional Schwarzschild black hole spacetimes}},
{\emph{Phys. Rev. D} {\bfseries 94}, 044052 (2016)},
[{{\ttfamily arXiv:1605.05263}}].

\bibitem{Konoplya:2003dd}
R.~A.~Konoplya,
\emph{{Gravitational quasinormal radiation of higher dimensional black holes}},
{\emph{Phys. Rev. D} {\bfseries 68}, 124017 (2003)},
[{{\ttfamily arXiv:hep-th/0309030}}].

\bibitem{Cardoso:2003vt}
V.~Cardoso, J.~P.~S.~Lemos, and S.~Yoshida,
\emph{{Quasinormal modes of Schwarzschild black holes in four-dimensions and higher dimensions}},
{\emph{Phys. Rev. D} {\bfseries 69}, 044004 (2004)},
[{{\ttfamily arXiv:gr-qc/0309112}}].

\bibitem{Seahra:2004fg}
S.~S.~Seahra, C.~Clarkson, and R.~Maartens,
\emph{{Detecting extra dimensions with gravity wave spectroscopy: the black string braneworld}},
{\emph{Phys. Rev. Lett.} {\bfseries 94}, 121302 (2005)},
[{{\ttfamily arXiv:gr-qc/0408032}}].

\bibitem{Kanti:2005xa}
P.~Kanti and R.~A.~Konoplya,
\emph{{Quasinormal modes of brane-localised standard model fields}},
{\emph{Phys. Rev. D} {\bfseries 73}, 044002 (2006)},
[{{\ttfamily  arXiv:hep-th/0512257}}].

\bibitem{Seahra:2006tm}
S.~S.~Seahra,
\emph{{Gravitational waves and cosmological braneworlds: A Characteristic evolution scheme}},
{\emph{Phys. Rev. D} {\bfseries 74}, 044010  (2006)},
[{{\ttfamily arXiv:hep-th/0602194}}].

\bibitem{Kanti:2006ua}
P.~Kanti, R.~A.~Konoplya, and A.~Zhidenko,
\emph{{Quasinormal Modes of Brane-Localised Standard Model Fields. II. Kerr Black Holes}},
{\emph{Phys. Rev. D} {\bfseries 74}, 064008 (2006)},
[{{\ttfamily  arXiv:gr-qc/0607048}}].

\bibitem{Ishihara:2008re}
H.~Ishihara, M.~Kimura, R.~A.~Konoplya, K.~Murata, J.~Soda, and A.~Zhidenko,
\emph{{Evolution of perturbations of squashed Kaluza-Klein black holes: escape from instability}},
{\emph{Phys. Rev. D} {\bfseries 77}, 084019 (2008)},
[{{\ttfamily  arXiv:0802.0655}}]. 

\bibitem{Chung:2015mna}
H.~Chung, L.~Randall, M.~J.~Rodriguez, and O.~Varela,
\emph{{Quasinormal ringing on the brane}},
{\emph{Class. Quant. Grav.} {\bfseries 33},  245013 (2016)},
[{{\ttfamily  arXiv:1508.02611}}].

\bibitem{Bronnikov:2019sbx}
K.~A.~Bronnikov and R.~A.~Konoplya,
\emph{{Echoes in brane worlds: ringing at a black hole--wormhole transition}},
{\emph{Phys. Rev. D} {\bfseries 101}, 064004 (2020)},
[{{\ttfamily  arXiv:1912.05315}}].

\bibitem{Dey:2020lhq}
R.~Dey, S.~Chakraborty, and N.~Afshordi,
\emph{{Echoes from braneworld black holes}},
{\emph{Phys. Rev. D} {\bfseries 101},  104014  (2020)},
[{{\ttfamily  arXiv:2001.01301}}].

\bibitem{Banerjee:2021aln}
I.~Banerjee, S.~Chakraborty, and S.~SenGupta,
\emph{{Looking for extra dimensions in the observed quasi-periodic oscillations of black holes}},
{\emph{JCAP} {\bfseries 09}, 037 (2021)},
[{{\ttfamily arXiv:2105.06636}}].

\bibitem{Mishra:2021waw}
A.~K.~Mishra, A.~Ghosh, and S.~Chakraborty,
\emph{{Constraining extra dimensions using observations of black hole quasinormal modes}},
{\emph{Eur. Phys. J. C} {\bfseries 82}, 820 (2022)},
[{{\ttfamily arXiv:2106.05558}}].

\bibitem{Lin:2022hus}
Z.-C.~Lin, H.~Yu, and Y.-X.~Liu,
\emph{{Shortcut in codimension-2 brane cosmology in light of GW170817}},
{\emph{Eur. Phys. J. C} {\bfseries 83}, 190 (2023)},
[{{\ttfamily arXiv:2202.04866}}].

\bibitem{Bohra:2023vls}
S.~S.~Bohra, S.~Sarkar, and A.~A.~Sen,
\emph{{Gravitational atoms in the braneworld scenario}},
{\emph{Phys. Rev. D} {\bfseries 109}, 104021 (2024)},
[{{\ttfamily arXiv:2312.07295}}].

\bibitem{Zinhailo:2024jzt}
A.~F.~Zinhailo,
\emph{{Exploring unique quasinormal modes of a massive scalar field in braneworld scenarios}},
{\emph{Phys. Lett. B} {\bfseries 853}, 138682 (2024)},
[{{\ttfamily arXiv:2403.06867}}].

%%%%%%%%%%%%%%%%%%%%%%%%%%%%%%%%%%%%%%%%%%%%%%%%%%%%%%%%%
% the QNMs of the thin and thick brane

\bibitem{Seahra:2005wk}
S.~S.~Seahra,
\emph{Ringing the Randall-Sundrum braneworld: Metastable gravity wave bound states},
{\emph{Phys. Rev. D} {\bfseries  72}, 066002 (2005)},
[{{\ttfamily arXiv:hep-th/0501175}}].

\bibitem{Seahra:2005iq}
S.~S.~Seahra,
\emph{Metastable massive gravitons from an infinite extra dimension},
{\emph{Int. J. Mod. Phys. D} {\bfseries 14},  2279 (2005)},
[{{\ttfamily arXiv:hep-th/0505196}}].

\bibitem{Tan:2022vfe}
Q.~Tan, W.-D.~Guo, and Y.-X.~Liu,
\emph{Sound from extra dimensions: Quasinormal modes of a thick brane},
{\emph{Phys. Rev. D} {\bfseries  106}, 044038 (2022)},
[{{\ttfamily arXiv:2205.05255}}].

\bibitem{Tan:2023cra}
Q.~Tan, W.-D.~Guo, Y.-P.~Zhang, and Y.-X.~Liu,
\emph{Characteristic modes of a thick brane: Resonances and quasinormal modes},
{\emph{Phys. Rev. D} {\bfseries 109},  024017 (2024)},
[{{\ttfamily arXiv:2304.09363}}].

\bibitem{Tan:2024url}
Q.~Tan, Y.~Zhong, and W.-D.~Guo,
\emph{Quasibound and quasinormal modes of a thick brane in Rastall gravity},
[{{\ttfamily arXiv:2404.11217}}].

%%%%%%%%%%%%%%%%%%%%%%%%%%%%%%%%%%%%%%%%%%%%%%%%%%%%%%%%%%%%%%%
% D brane

\bibitem{Polchinski:1995mt}
J.~Polchinski,
\emph{Dirichlet Branes and Ramond-Ramond charges},
{\emph{Phys. Rev. Lett.} {\bfseries  75},  4724 (1995)},
[{{\ttfamily arXiv:hep-th/9510017}}].

%%%%%%%%%%%%%%%%%%%%%%%%%%%%%%%%%%%%%%%%%%%%%%%%%%%%%%%%%%%%%%%
% thin brane model 

\bibitem{Antoniadis:1990ew}
I.~Antoniadis, 
\emph{{A Possible new dimension at a few TeV}},
{\emph{Phys. Lett. B} {\bfseries 246}, 377 (1990)}.

\bibitem{Arkani-Hamed:1998jmv}
N.~Arkani-Hamed, S.~Dimopoulos, and G.~R.~Dvali,
\emph{The Hierarchy problem and new dimensions at a millimeter},
{\emph{Phys. Lett. B} {\bfseries  429},  263 (1998)},
[{{\ttfamily arXiv:hep-ph/9803315}}].

\bibitem{Antoniadis:1998ig}
I.~Antoniadis, N.~Arkani-Hamed, S.~Dimopoulos, and G.~R.~Dvali,
\emph{New dimensions at a millimeter to a Fermi and superstrings at a TeV},
{\emph{Phys. Lett. B} {\bfseries  436},  257 (1998)},
[{{\ttfamily arXiv:hep-ph/9804398}}].

\bibitem{Randall:1999ee}
L.~Randall and R.~Sundrum, 
\emph{{A Large mass hierarchy from a small extra dimension}},
{\emph{Phys. Rev. Lett.} {\bfseries 83},  3370 (1999)},
[{{\ttfamily arXiv:hep-ph/9905221}}].

\bibitem{Randall:1999vf}
L.~Randall and R.~Sundrum, 
\emph{{An Alternative to compactification}},
{\emph{Phys. Rev. Lett.} {\bfseries 83},  4690 (1999)},
[{{\ttfamily arXiv:hep-th/9906064}}].

%%%%%%%%%%%%%%%%%%%%%%%%%%%%%%%%%%%%%%%%%%%%%%%%%%%%%%
% 52 mu m

\bibitem{Tu:2007zz}
L.-C.~Tu, S.-G.~Guan, J.~Luo, C.-G.~Shao, and L.-X.~Liu,
\emph{{Null Test of Newtonian Inverse-Square Law at Submillimeter Range with a Dual-Modulation Torsion Pendulum}},
{\emph{Phys. Rev. Lett.} {\bfseries 98}, 201101 (2007)}.

\bibitem{Yang:2012zzb}
S.-Q.~Yang, B.-F.~Zhan, Q.-L.~Wang, C.-G.~Shao, L.-C.~Tu, W.-H.~Tan, and J.~Luo,
\emph{{Test of the Gravitational Inverse Square Law at Millimeter Ranges}},
{\emph{Phys. Rev. Lett.} {\bfseries 108}, 081101 (2012)}.

\bibitem{Tan:2016vwu}
W.-H.~Tan, S.-Q.~Yang, C.-G.~Shao, J.~Li, A.-B.~Du, B.-F.~Zhan, Q.-L.~Wang, P.-S.~Luo, L.-C.~Tu, and J.~Luo,
\emph{{New Test of the Gravitational Inverse-Square Law at the Submillimeter Range with Dual Modulation and Compensation}},
{\emph{Phys. Rev. Lett.} {\bfseries 116}, 131101 (2016)}.

\bibitem{Tan:2020vpf}
W.-H.~Tan, A.-B.~Du, and W.-C.~Dong, \textit{et al.}
\emph{{Improvement for Testing the Gravitational Inverse-Square Law at the Submillimeter Range}},
{\emph{Phys. Rev. Lett.} {\bfseries 124}, 051301 (2020)}.

\bibitem{Lee:2020zjt}
J.~G.~Lee, E.~G.~Adelberger, T.~S.~Cook, S.~M.~Fleischer, and B.~R.~Heckel,
\emph{{New Test of the Gravitational $1/r^2$ Law at Separations down to 52 $\mu$m}},
{\emph{Phys. Rev. Lett.} {\bfseries 124},  101101 (2020)},
[{{\ttfamily arXiv:2002.11761}}].

\bibitem{Du:2022veu}
A.-B.~Du, W.-H.~Tan, W.-C.~Dong, H.~Huang, L.~Zhu, Y.-J.~Tan, C.-G.~Shao, S.-Q.~Yang, and J.~Luo,
\emph{{A new design for testing the gravitational inverse-square law at the sub-millimeter range with a 32-fold symmetric attractor}},
{\emph{Class. Quant. Grav.} {\bfseries 39}, 105008 (2022)}.

%%%%%%%%%%%%%%%%%%%%%%%%%%%%%%%%%%%%%%%%%%%%%%%%%%%%%%
% domain wall model article

\bibitem{Akama:1982jy}
K.~Akama,
\emph{{An Early Proposal of `Brane World'}},
{\emph{Lect. Notes Phys.} {\bfseries 176}, 267 (1982)},
[{{\ttfamily arXiv:hep-th/0001113}}].

\bibitem{Rubakov:1983bb}
V.~A.~Rubakov and M.~E.~Shaposhnikov,
\emph{{Do We Live Inside a Domain Wall?}}
{\emph{Phys. Lett. B} {\bfseries 125}, 136 (1983)}.

%%%%%%%%%%%%%%%%%%%%%%%%%%%%%%%%%%%%%%%%%%%%%%%%%%%%%%%%%%%%%%%%%%%%%%%%%%%%%%%%
% the first article of thick brane

\bibitem{DeWolfe:1999cp}
O.~DeWolfe, D.~Z.~Freedman, S.~S.~Gubser, and A.~Karch,
\emph{{Modeling the fifth-dimension with scalars and gravity}},
{\emph{Phys. Rev. D} {\bfseries 62}, 046008 (2000)},
[{{\ttfamily arXiv:hep-th/9909134}}].

\bibitem{Gremm:1999pj}
M.~Gremm,
\emph{{Four-dimensional gravity on a thick domain wall}},
{\emph{Phys. Lett. B} {\bfseries 478}, 434 (2000)},
[{{\ttfamily arXiv:hep-th/9912060}}].

\bibitem{Csaki:2000fc}
C.~Csaki, J.~Erlich, T.~J.~Hollowood, and Y.~Shirman,
\emph{{Universal aspects of gravity localized on thick branes}},
{\emph{Nucl. Phys. B} {\bfseries 581}, 309 (2000)},
[{{\ttfamily arXiv:hep-th/0001033}}].

%%%%%%%%%%%%%%%%%%%%%%%%%%%%%%%%%%%%%%%%%%%%%%%%%%%%%%%%%%%%%%%%%%%
% some thick brane solution 
\bibitem{Wang:2002pka}
A.-Z.~Wang, 
\emph{{Thick de Sitter 3 branes, dynamic black holes and localization of gravity}},
{\emph{Phys. Rev. D} {\bfseries 66}, 024024 (2002)},
[{{\ttfamily arXiv:hep-th/0201051}}].

\bibitem{Dzhunushaliev:2010fqo}
V.~Dzhunushaliev and V.~Folomeev,
\emph{{Spinor brane}},
{\emph{Gen. Rel. Grav.} {\bfseries 43}, 1253 (2011)},
[{{\ttfamily arXiv:0909.2741}}].

\bibitem{Dzhunushaliev:2011mm}
V.~Dzhunushaliev and V.~Folomeev,
\emph{{Thick brane solutions supported by two spinor fields}},
{\emph{Gen. Rel. Grav.} {\bfseries 44}, 253 (2012)},
[{{\ttfamily arXiv:1104.2733}}].

\bibitem{Guo:2011wr}
H.~Guo, Y.-X.~Liu, Z.-H.~Zhao, and F.-W.~Chen,
\emph{{Thick branes with a nonminimally coupled bulk-scalar field}},
{\emph{Phys. Rev. D} {\bfseries 85}, 124033 (2012)},
[{{\ttfamily arXiv:1106.5216}}].

\bibitem{Liu:2012gv}
Y.-X.~Liu, F.-W.~Chen, H.~Guo, and X.-N.~Zhou,
\emph{{Nonminimal Coupling Branes}},
{\emph{JHEP} {\bfseries 05}, 108 (2012)},
[{{\ttfamily arXiv:1205.0210}}].

\bibitem{German:2012rv}
G.~German, A.~Herrera-Aguilar, D.~Malagon-Morejon, R.~R.~Mora-Luna, and R.~da Rocha,
\emph{{A de Sitter tachyon thick braneworld and gravity localization}},
{\emph{JCAP} {\bfseries 02}, 035 (2013)},
[{{\ttfamily arXiv:1210.0721}}].

%%%%%%%%%%%%%%%%%%%%%%%%%%%%%%%%%%%%%%%%%%%%%%%%%%%%%%%%%%%%%%%%%%%
% localization of gravity and matter fields 
\bibitem{Gregory:2000jc}
R.~Gregory, V.~A.~Rubakov, and S.~M.~Sibiryakov,
\emph{{Opening up extra dimensions at ultra large scales}},
{\emph{Phys. Rev. Lett.} {\bfseries 84}, 5928 (2000)},
[{{\ttfamily arXiv:hep-th/0002072}}].

\bibitem{Melfo:2006hh}
A.~Melfo, N.~Pantoja, and J.~D. Tempo, 
\emph{{Fermion localization on thick branes}},
{\emph{Phys. Rev. D} {\bfseries 73}, 044033 (2006)},
[{{\ttfamily arXiv:hep-th/0601161}}].

\bibitem{Almeida:2009jc}
C.~A. Almeida, R.~Casana, M.~M. Ferreira, and A.~R. Gomes, 
\emph{{Fermion localization and resonances on two-field thick branes}},  
{\emph{Phys. Rev. D} {\bfseries 79}, 125022 (2009)}, 
[{{\ttfamily arXiv:0901.3543}}].

\bibitem{Liu:2009ve}
Y.-X.~Liu, J.~Yang, Z.-H.~Zhao, C.-E.~Fu, and Y.-S.~Duan,
\emph{{Fermion Localization and Resonances on A de Sitter Thick Brane}},
{\emph{Phys. Rev. D} {\bfseries 80}, 065019 (2009)},
[{{\ttfamily arXiv:0904.1785}}].

\bibitem{Liu:2009dw}
Y.-X.~Liu, C.-E.~Fu, L.~Zhao, and Y.-S.~Duan,
\emph{{Localization and Mass Spectra of Fermions on Symmetric and Asymmetric Thick Branes}},
{\emph{Phys. Rev. D} {\bfseries 80}, 065020 (2009)},
[{{\ttfamily arXiv:0907.0910}}].

\bibitem{Liu:2009uca}
Y.-X.~Liu, H.~Guo, C.-E.~Fu, and J.-R.~Ren,
\emph{{Localization of Matters on Anti-de Sitter Thick Branes}},
{\emph{JHEP} {\bfseries 02}, 080 (2010)},
[{{\ttfamily arXiv:0907.4424}}].

\bibitem{Zhao:2009ja}
Z.-H.~Zhao, Y.-X.~Liu, and H.-T.~Li,
\emph{{Fermion localization on asymmetric two-field thick branes}},  
{\emph{Class. Quantum Gravity} {\bfseries 27}, 185001 (2010)}, 
[{{\ttfamily arXiv:0911.2572}}].

\bibitem{Liu:2011wi}
Y.-X.~Liu, Y.~Zhong, Z.-H.~Zhao, and H.-T.~Li, 
\emph{{Domain wall brane in squared curvature gravity}},  
{\emph{J. High Energy Phys.} {\bfseries 06}, 135 (2011)}, 
[{{\ttfamily arXiv:1104.3188}}].

\bibitem{Xie:2015dva}
Q.-Y.~Xie, H.~Guo, Z.-H.~Zhao, Y.-Z.~Du, and Y.-P.~Zhang, 
\emph{{Spectrum structure of a fermion on Bloch branes with two scalar-fermion couplings}}, 
{\emph{Class. Quantum Gravity} {\bfseries 34}, 055007 (2017)}, 
[{{\ttfamily arXiv:1510.03345}}].

\bibitem{Gu:2016nyo}
B.-M.~Gu, Y.-P.~Zhang, H.~Yu, and Y.-X.~Liu, 
\emph{{Full linear perturbations and localization of gravity on $f(R, T)$ brane}},  
{\emph{Eur. Phys. J. C} {\bfseries 77}, 115 (2017)}, 
[{{\ttfamily arXiv:1606.07169}}].

\bibitem{Zhong:2016iko}
Y.~Zhong and Y.-X.~Liu,
\emph{{Linearization of a warped $f(R)$ theory in the higher-order frame}},
{\emph{Phys. Rev. D} {\bfseries 95}, 104060 (2017)},
[{{\ttfamily arXiv:1611.08237}}].

\bibitem{Zhou:2017xaq}
X.-N.~Zhou, Y.-Z.~Du, H.~Yu, and Y.-X.~Liu,
\emph{{Localization of gravitino field on $f(R)$-thick branes}},
{\emph{Sci. China Physics, Mech. Astron.} {\bfseries 61}, 110411 (2018)},
[{{\ttfamily arXiv:1703.10805}}].

\bibitem{Xie:2021ayr}
Q.-Y.~Xie, Q.-M.~Fu, T.-T.~Sui, L.~Zhao, and Y.~Zhong,
\emph{{First-Order Formalism and Thick Branes in Mimetic Gravity}},
{\emph{Symmetry} {\bfseries 13}, 1345 (2021)},
[{{\ttfamily arXiv:2102.10251}}].

\bibitem{Xu:2022xxd}
N.~Xu, J.~Chen, Y.-P.~Zhang, and Y.-X.~Liu,
\emph{{Multikink brane in Gauss-Bonnet gravity and its stability}},
{\emph{Phys. Rev. D} {\bfseries 107}, 124011 (2023)},
[{{\ttfamily arXiv:2201.10282}}].

%%%%%%%%%%%%%%%%%%%%%%%%%%%%%%%%%%%%%%%%%%%%%%%%%%%%%%%%%%%%%%%%%%%%%%%
% review article

\bibitem{Liu:2017gcn}
Y.-X.~Liu,
\emph{{Introduction to Extra Dimensions and Thick Braneworlds}},
[{{\ttfamily arXiv:1707.08541}}].

\bibitem{Dzhunushaliev:2009va}
V.~Dzhunushaliev, V.~Folomeev, and M.~Minamitsuji,
\emph{{Thick brane solutions}},
{\emph{Rept. Prog. Phys.} {\bfseries 73}, 066901 (2010)},
[{{\ttfamily arXiv:0904.1775}}].

\bibitem{Ahluwalia:2022ttu}
D.~V.~Ahluwalia, J.~M.~H.~da Silva, C.~Y.~Lee, Y.-X.~Liu, S.~H.~Pereira, and M.~M.~Sorkhi,
\emph{{Mass dimension one fermions: Constructing darkness}},
{\emph{Phys. Rept.} {\bfseries 967}, 1 (2022)},
[{{\ttfamily arXiv:2205.04754}}].
%%%%%%%%%%%%%%%%%%%%%%%%%%%%%%%%%%%%%%%%%%%%%%%%%%%%%%
% superpotential
\bibitem{Bazeia:2008zx}
D.~Bazeia, A.~R.~Gomes, L.~Losano, and R.~Menezes,
\emph{{Braneworld Models of Scalar Fields with Generalized Dynamics}},
{\emph{Phys. Lett. B} {\bfseries 671}, 402 (2009)},
[{{\ttfamily arXiv:0808.1815}}].

\bibitem{Zhong:2013xga}
Y.~Zhong, Y.-X.~Liu, and Z.-H.~Zhao,
\emph{{Nonperturbative procedure for stable K-brane}},
{\emph{Phys. Rev. D} {\bfseries 89}, 104034 (2014)},
[{{\ttfamily arXiv:1401.0004}}].

%%%%%%%%%%%%%%%%%%%%%%%%%%%%%%%%%%%%%%%%%%%%%%%%%%%
% select the warp factor article

\bibitem{Gremm:2000dj}
M.~Gremm,
\emph{{Thick domain walls and singular spaces}},
{\emph{Phys. Rev. D} {\bfseries 62}, 044017 (2000)},
[{{\ttfamily arXiv:hep-th/0002040}}].

\bibitem{Barbosa-Cendejas:2007ucz}
N.~Barbosa-Cendejas, A.~Herrera-Aguilar, U.~Nucamendi, I.~Quiros, and K.~Kanakoglou,
\emph{{Mass hierarchy, mass gap and corrections to Newton's law on thick branes with Poincare symmetry}},
{\emph{Gen. Rel. Grav.} {\bfseries 46}, 1631 (2014)},
[{{\ttfamily arXiv:0712.3098}}].

%%%%%%%%%%%%%%%%%%%%%%%%%%%%%%%%%%%%%%%%%%%%%%%%%%%%%%%%%%%%%%%%%%
% scalar perturbation

\bibitem{Giovannini:2001fh}
M.~Giovannini,
\emph{{Gauge invariant fluctuations of scalar branes}},
{\emph{Phys. Rev. D} {\bfseries 64}, 064023 (2001)},
[{{\ttfamily arXiv:hep-th/0106041}}].

\bibitem{Giovannini:2001xg}
M.~Giovannini,
\emph{{Localization of metric fluctuations on scalar branes}},
{\emph{Phys. Rev. D} {\bfseries 65}, 064008 (2002)},
[{{\ttfamily arXiv:hep-th/0106131}}].

\bibitem{Kobayashi:2001jd}
S.~Kobayashi, K.~Koyama, and J.~Soda,
\emph{{Thick brane worlds and their stability}},
{\emph{Phys. Rev. D} {\bfseries 65}, 064014 (2002)},
[{{\ttfamily arXiv:hep-th/0107025}}].

%%%%%%%%%%%%%%%%%%%%%%%%%%%%%%%%%%%%%%%%%%%%%%%%%%%%%%%%%%%%%%%%%%
% PT potential

\bibitem{D_az_1999}
J.~I.~Díaz, J.~Negro, L.~M.~Nieto, and O.~Rosas-Ortiz,
\emph{{The supersymmetric modified Pöschl-Teller and delta well potentials}}
{\emph{Journal of Physics A: Mathematical and General} {\bfseries 32}, 8447 (1999)}.

%%%%%%%%%%%%%%%%%%%%%%%%%%%%%%%%%%%%%%%%%%%%%%%%%%%%%%%%%%%%%%%%%%
% dS space QNMs

\bibitem{Natario:2004jd}
J.~Natario and R.~Schiappa,
\emph{{On the classification of asymptotic quasinormal frequencies for d-dimensional black holes and quantum gravity}},
{\emph{Adv. Theor. Math. Phys.} {\bfseries 8}, 1001 (2004)},
[{{\ttfamily arXiv:hep-th/0411267}}].

%%%%%%%%%%%%%%%%%%%%%%%%%%%%%%%%%%%%%%%%%%%%%%%%%%%%%%%%%%%%%%%%%%
% continued fraction method

\bibitem{Leaver:1985ax}
E.~W.~Leaver,
\emph{{An Analytic representation for the quasinormal modes of Kerr black holes}},
{\emph{Proc. Roy. Soc. Lond. A} {\bfseries 402}, 285 (1985)}.

\bibitem{Leaver:1990zz}
E.~W.~Leaver, 
\emph{{Quasinormal modes of Reissner-Nordstrom black holes}},
{\emph{Phys. Rev. D} {\bfseries 41}, 2986 (1990)}.

%%%%%%%%%%%%%%%%%%%%%%%%%%%%%%%%%%%%%%%%%%%%%%%%%%%%%%%%%%%%%%%%%%
% WKB method

\bibitem{Schutz:1985km}
B.~F.~Schutz and C.~M.~Will,
\emph{Black hole normal modes: a semianalytic approach},
{\emph{Astrophys. J. Lett.} {\bfseries  291}, L33 (1985)}.

%%%%%%%%%%%%%%%%%%%%%%%%%%%%%%%%%%%%%%%%%%%%%%%%%%%%%%%%%%%%%%%%%%
% AIM method

\bibitem{Cho:2011sf}
H.-T.~Cho, A.~S.~Cornell, J.~Doukas, T.-R.~Huang, and W.~Naylor,
\emph{A New Approach to Black Hole Quasinormal Modes: A Review of the Asymptotic Iteration Method},
{\emph{Adv. Math. Phys.} {\bfseries  2012}, 281705 (2012)},
[{{\ttfamily arXiv:1111.5024}}].

%%%%%%%%%%%%%%%%%%%%%%%%%%%%%%%%%%%%%%%%%%%%%%%%%%%%%%%%%%%%%%%%%%
% shooting method

\bibitem{Pani:2013pma}
P.~Pani,
\emph{Advanced Methods in Black-Hole Perturbation Theory},
{\emph{Int. J. Mod. Phys. A} {\bfseries  28}, 1340018 (2013)},
[{{\ttfamily arXiv:1305.6759}}].

%%%%%%%%%%%%%%%%%%%%%%%%%%%%%%%%%%%%%%%%%%%%%%%%%%%%%%%%%%%%%%%%%%
% Supersymmetry potential

\bibitem{Cooper:1994eh}
F.~Cooper, A.~Khare, and U.~Sukhatme,
\emph{{Supersymmetry and quantum mechanics}},
{\emph{Phys. Rept.} {\bfseries 251},  267 (1995)},
[{{\ttfamily arXiv:hep-th/9405029}}].

%%%%%%%%%%%%%%%%%%%%%%%%%%%%%%%%%%%%%%%%%%%%%%%%%%%%%%%%%%%%%%%%%%
% numerical evolution

\bibitem{Gundlach:1993tp}
C.~Gundlach, R.~H.~Price and J.~Pullin,
\emph{{Late time behavior of stellar collapse and explosions: 1. Linearized perturbations}},
{\emph{Phys. Rev. D} {\bfseries 49},  883 (1994)},
[{{\ttfamily arXiv:gr-qc/9307009}}].

%%%%%%%%%%%%%%%%%%%%%%%%%%%%%%%%%%%%%%%%%%%%%%%%%%%%%%%%%%%%%%%%%%
% Newtonian potential law limit 

\bibitem{Sui:2020atb}
T.-T.~Sui, Y.-P.~Zhang, B.-M.~Gu, and Y.-X.~Liu,
\emph{{Fundamental energy scale of the thick brane in mimetic gravity}},
{\emph{Eur. Phys. J. C} {\bfseries 81},  980 (2021)},
[{{\ttfamily arXiv:2005.08438}}].

%%%%%%%%%%%%%%%%%%%%%%%%%%%%%%%%%%%%%%%%%%%%%%%%%%%%%%%%%%%%%%%%%%
% High-frequency gravitational wave

\bibitem{Li:2017jcz}
F.-Y.~Li, H.~Wen, Z.-Y.~Fang, D.~Li, and T.-J.~Zhang,
\emph{{Electromagnetic response to high-frequency gravitational waves having additional polarization states: distinguishing and probing tensor-mode, vector-mode and scalar-mode gravitons}},
{\emph{Eur. Phys. J. C} {\bfseries 80},  879 (2020)},
[{{\ttfamily arXiv:1712.00766}}].

\bibitem{Zhao:2021zlr}
Z.-C.~Zhao and Z.-J.~Cao,
\emph{{Stochastic gravitational wave background due to gravitational wave memory}},
{\emph{Sci. China Phys. Mech. Astron.} {\bfseries 65},  119511 (2022)},
[{{\ttfamily arXiv:2111.13883}}].

%%%%%%%%%%%%%%%%%%%%%%%%%%%%%%%%%%%%%%%%%%%%%%%%%%%%%%%%%%%%%%%%%%
% pure imaginary QNMs of BH

\bibitem{Cook:2016fge}
G.~B.~Cook and M.~Zalutskiy,
\emph{{Purely imaginary quasinormal modes of the Kerr geometry}},
{\emph{Class. Quant. Grav.} {\bfseries 33},  245008 (2016)},
[{{\ttfamily arXiv:1603.09710}}].

\bibitem{Konoplya:2023fmh}
R.~A.~Konoplya and A.~Zhidenko,
\emph{{Asymptotic tails of massive gravitons in light of pulsar timing array observations}},
{\emph{Phys. Lett. B} {\bfseries 853},  138685 (2024)},
[{{\ttfamily arXiv:2307.01110}}].

\bibitem{NANOGrav:2023gor}
G.~Agazie \textit{et al.} [NANOGrav],
\emph{{The NANOGrav 15 yr Data Set: Evidence for a Gravitational-wave Background}},
{\emph{Astrophys. J. Lett.} {\bfseries 951},  L8 (2023)},
[{{\ttfamily arXiv:2306.16213}}].

\bibitem{Koyama:2004cf}
K.~Koyama,
\emph{{Late time behavior of cosmological perturbations in a single brane model}},
{\emph{JCAP} {\bfseries 09},  010 (2004)},
[{{\ttfamily arXiv:astro-ph/0407263}}].

\bibitem{Caprini:2018mtu}
C.~Caprini and D.~G.~Figueroa,
\emph{{Cosmological Backgrounds of Gravitational Waves}},
{\emph{Class. Quant. Grav.} {\bfseries 35},  163001 (2018)},
[{{\ttfamily arXiv:1801.04268}}].

\end{thebibliography}
\end{document}